\documentclass[useAMS,usenatbib]{mnras}
\usepackage{amsmath,fleqn,graphicx,amssymb,xcolor}
\usepackage{multirow}
\renewcommand{\[}{\begin{equation}}
\renewcommand{\]}{\end{equation}}
\def\aFe{$[\alpha/\hbox{Fe}]$}
\def\Grp{G_{\rm rp}} \def\Gbp{G_{\rm bp}}
\def\yr{\,{\rm yr}}
\def\AGAMA{{\it AGAMA}}

\let\boldgrk=\gkvecten
\let\boldgrksc=\gkvecseven

\def\gkthing#1{{\mathchoice%
	{\hbox{{\boldgrk\char#1}}}
	{\hbox{{\boldgrk\char#1}}}
	{\hbox{{\boldgrksc\char#1}}}
	{\hbox{{\boldgrksc\char#1}}}}}

\def\vtheta{\gkthing{18}}

{\newif\ifnotend
\notendtrue
\def\veclist{ABCDEFGHIJKLMNOPQRSTUVWXYZabcdefghijklmnopqrstuvwxyz.}
\def\top#1#2.{#1}
\def\tail#1#2.{#2.}
\loop\expandafter\xdef\csname v\expandafter\top\veclist\endcsname%
{{\noexpand\bf\expandafter\top\veclist}}
\edef\veclist{\expandafter\tail\veclist}
\if\veclist.\notendfalse\fi\ifnotend\repeat}
{\newif\ifnotend
\notendtrue
\def\callist{ABCDEFGHIJKLMNOPQRSTUVWXYZ.}
\def\top#1#2.{#1}
\def\tail#1#2.{#2.}
\loop\expandafter\xdef\csname c\expandafter\top\callist\endcsname%
{{\noexpand\cal\expandafter\top\callist}}
\edef\callist{\expandafter\tail\callist}
\if\callist.\notendfalse\fi\ifnotend\repeat}

\def\d{{\rm d}}

\def\mnras{MNRAS}
\def\araa{AnnRA\&A}
\def\apj{ApJ}
\def\apjs{ApJS}
\def\apjl{ApJL}
\def\nat{Nat}
\def\aj{AJ}
\def\aap{A\&A}

\def\pasp{PASP}

\def\Gyr{\,\mathrm{Gyr}}
\def\Myr{\,\mathrm{Myr}}
\def\kpc{\,\mathrm{kpc}}

\def\kms{\,\mathrm{km\,s}^{-1}}

\def\msun{\,{\rm M}_\odot}

\def\pc{\,\mathrm{pc}}
\def\e{\mathrm{e}}

\def\fracj#1#2{{\textstyle{#1\over#2}}}

\def\eqrf#1{(\ref{#1})}

\title[Young disc]{Our Galaxy's  youngest disc}

\author[Chengdong Li \& James Binney]{
  Chengdong Li$^1$\thanks{E-mail: chenglong.li@physics.ox.ac.uk} 
\& James Binney$^1$\thanks{E-mail:
  binney@physics.ox.ac.uk}\\   
  $^1$Rudolf Peierls Centre for Theoretical Physics, Clarendon Laboratory,
  Parks Road, Oxford, OX1 3PU, UK
}

\date{}

\pubyear{}

\begin{document}
\label{firstpage}
\pagerange{\pageref{firstpage}--\pageref{lastpage}}
\maketitle

\begin{abstract}
We investigate the structure of our Galaxy's young stellar disc by fitting
the distribution functions (DFs) of a new family to five-dimensional Gaia
data for a sample of $47\,000$ OB stars. Tests of the fitting procedure show
that the young disc's DF would be strongly constrained by Gaia data if the
distribution of Galactic dust were accurately known. The DF that best fits
the real data accurately predicts the kinematics of stars at their observed
locations, but it predicts the spatial distribution of stars poorly, almost
certainly on account of errors in the best-available dust map. We argue that
dust models could be greatly improved by modifying the dust model until the
spatial distribution of stars predicted by a DF agreed with the data. The
surface density of OB stars is predicted to peak at $R\simeq5.5\kpc$,
slightly outside the reported peak in the surface density of molecular gas;
we suggest that the latter radius may have been under-estimated through the
use of poor kinematic distances. The
velocity distributions predicted by the best-fit DF for stars with measured
line-of-sight velocities $v_\parallel$ reveal that the outer disc is
disturbed at the level of $10\kms$ in agreement with earlier studies, and
that the measured values of $v_\parallel$ have significant contributions from
the orbital velocities of binaries. Hence the outer disc is colder than it
is sometimes reported  to be. 
\end{abstract}




\section{Introduction}\label{sec:intro}

Over the last two decades our Galaxy has been the target of intense
observational activity not only on account of its interest as our home but
also because of its cosmological significance: it is a uniquely accessible
example of the type of galaxy that currently dominates the cosmic
star-formation rate. Data from massive photometric \citep{Schmidt2005,Skrutskie2006,Kaiser2010}
spectroscopic \citep{Majewski2017,Steinmetz2006,DeSilva2015} and astrometric
\citep{GaiaCollaboration2016} surveys are now in hand and we need to
synthesise these data into a coherent physical picture of our archetypal
Galaxy.

The stellar distribution of our Galaxy is dominated by the disc. Over the
last half century it has become conventional to decompose the disc into
several components. On the largest scale, \cite{Gilmore1983} pointed out that
the disc is split into thin and thick components. More recently, in light of
spectroscopic data rather than star-counts, the view is gaining ground that
the fundamental division is between a disc of very old stars with \aFe\
larger than the Sun and a continuously-forming disc of stars with `normal'
values of \aFe\ \citep{Hayden2015,BlandHawthorn2019}.  The former, very old disc has
a scale height $z_0>0.7\kpc$ at all radii, while the latter disc has a much
smaller $(\sim0.3\kpc$) scale height, except possibly beyond the solar radius
$R_0$, where it probably flares. The thin, or `$\alpha$-normal'
disc is naturally divided into sub-discs comprising stars of similar ages
because the $\alpha$-normal disc has formed continuously over $8-10\Gyr$ and
older stellar cohorts now have larger random velocities and vertical
scale-heights. It is also thought that older cohorts have smaller radial
scale-lengths.

Recently \citep{Li2022} we investigated the structure of our Galaxy's stellar
halo by modelling the distribution of stars identified as RR-Lyrae variables
in data from the Pan-STARRS survey \citep{Kaiser2010}. In that paper we
developed a technique for modelling a Galactic component using the
five-dimensional data for stars that is available in enormous quantities from
the Gaia mission \citep[][and references therein]{GaiaCollaboration2021b}. In
this paper we apply this technique to objects that are likely OB stars, which
may be considered tracers of the youngest cohort in the disc. The structure
of this component is intrinsically of great interest, but modelling it forces
one to engage with problems that do not arise when modelling the RR-Lyrae
population.  

We model a component by fitting to the data a parametrised distribution
function (DF) $f(\vJ)$ that is a function of the action integrals within a
given model of the Galaxy's gravitational potential. This potential itself
emerges from fitting DFs for several components, both stellar and dark, to
six-dimensional Gaia data \citep[][hereafter BV2022]{BinneyVasiliev2022}. The
potential is that co-operatively generated by the DFs of dark matter and all
the major stellar components together with a gas disc of pre-determined
structure. The models are constructed and analysed using the \AGAMA\ software
library \citep{Vasiliev2019}.

When we lack data for one of the six phase-space coordinates, as we do for
nearly all the $\sim1.3$ billion stars monitored by Gaia, fitting a DF to
data is more computationally demanding because one has marginalise over the
unknown line-of-sight velocity $v_\parallel$ of each star. Against this
significant disadvantage may be set two substantial advantages: (i) in the
Early Third Data Release (EDR3) from Gaia \citep{GaiaCollaboration2021b} contains
$\sim1.3\times10^9$ stars whereas Gaia's Radial Velocity Sample RVS
\citep{GaiaCollaboration2018b} contains only $\sim7\times10^6$ stars, and (ii) while the
selection function (SF) relevant to five-dimensional data is complex, it is
much better known than that of the RVS \citep{Boubert2021,Everall2021}. Hence, when
using five-dimensional data significance can be attached to the density of
stars  in real space in a way that is not possible when using
six-dimensional data: in that case only densities in velocity space are
significant.

The Galaxy's young stellar disc is perhaps the  hardest component  to observe
because it is largely confined to a thin layer around the Galactic plane and
is therefore heavily obscured, even fairly close to the Sun. We will find
that our limited knowledge of the distribution of dust through the disc severely
limits our ability to determine the structure of the young disc.

This paper is structured as follows. Section \ref{sec:sample} explains how we
extracted a sample of OB stars from Gaia EDR3. In Section
\ref{sec:methodology} we outline the approach to modelling five-dimensional
data that was explained in \cite{Li2022}; Appendix~\ref{app:LiB} gives
greater detail. Section \ref{sec:DF} specifies the structure of the DF
$f(\vJ)$ that we fit to our tracers of the young disc.  Section
\ref{sec:potential} outlines the self-consistent Galaxy model and the dust
model in which we place models of the young disc.  In section \ref{sec:tests}
we use mock data to discover how accurately we can determine the structure of
the young disc from a sample of OB stars.  In Section \ref{sec:fits} we fit
DFs to our sample of OB stars and conclude that the limited spatial extent of
the sample, combined with defects in dust models, severely limit our ability
to determine the large-scale structure of the young disc. In Section
\ref{sec:quality} we investigate how well the best-fitting models account for
the data. We find that while the models reproduce the observed kinematics
well, defects of the dust models prevent the DFs reproducing the distribution
of stars on the sky. In Section \ref{sec:LAMOST} we compare the predictions
of our models to six-dimensional data from the LAMOST spectroscopic survey.
We confirm the conclusion of \cite{Eilers2020} that stars at $R\sim12\kpc$ in
the anticentre direction are moving systematically outwards as a consequence
of large-scale disturbance of the disc. We argue that the orbital velocities
of binaries make
a significant contribution to the measured line-of-sight velocities of OB
stars, which is  why several studies have found a puzzling increase
with Galactocentric radius in the in-plane velocity dispersion of young
stars. Section \ref{sec:conclude} sums up and identifies useful next steps.

\section{A sample of OB stars}\label{sec:sample}

We selected OB stars from the intersection of three catalogues: Gaia's
\citep{GaiaCollaboration2016} early third data release
\citep{GaiaCollaboration2021b}, the 2MASS infrared catalogue
\citep{Skrutskie2006} and the the Starhorse catalogue
\citep{Anders2019,Anders2021}\footnote{\textit{Gaia}@AIP
Services at https://gaia.aip.de/}, which provides Bayesian fits of
distances and astrophysical parameters to all stars in EDR3 brighter than
$G=18.5$. EDR3, which was released in December of 2020, contains precision
astrometry and photometry for 1.5 billion stars. Parallaxes and proper
motions have typical uncertainties of $0.07\,$mas and $0.07\,$mas$\yr^{-1}$
at $G = 17\,$mag and $0.5\,$mas and $0.5\,$mas$\yr^{-1}$ at $G = 20\,$mag
\citep{GaiaCollaboration2021b}. EDR3 gives magnitudes in red and blue
passbands $\Grp$ and $\Gbp$ in addition to the broad-band $G$ magnitudes. 

We start by selecting stars that satisfy three 
basic criteria \citep{GaiaCollaboration2018b}:
\begin{equation}
    \begin{aligned}
    &\varpi/\epsilon_{\varpi}\,>\,5,\\
    &(\Gbp-\Grp)_{0}=(\Gbp-\Grp)\,-\,E(\Gbp-\Grp)\,<\,0,\\
    &M_G=G-5\log(s/10\pc)-A_{G}\,<\,2,
    \end{aligned}
    \label{eq:sfgaia}
\end{equation}
where $A_G$ and $E(\Gbp-\Grp)$ are the star's extinction and colour
from  the Starhorse catalogue. In order to avoid contamination
by red giants and red clump stars, 2MASS photometry  \citep{Skrutskie2006}
is now used to make a further selection. We consider only stars with
photometric flag AAA that are blue enough to satisfy
\begin{equation}
    \begin{aligned}
    &J-H<0.14(G-K_{s})+0.02,\\
    &J-K_{s}<0.23(G-K_{s}).
    \end{aligned}
    \label{eq:2mass}
\end{equation}
Then, we exclude A stars by restricting the sample to stars  with
$T_{\rm eff}>10\,000\,$K in the Starhorse catalogue.

\begin{figure}
	\includegraphics[width=\columnwidth]{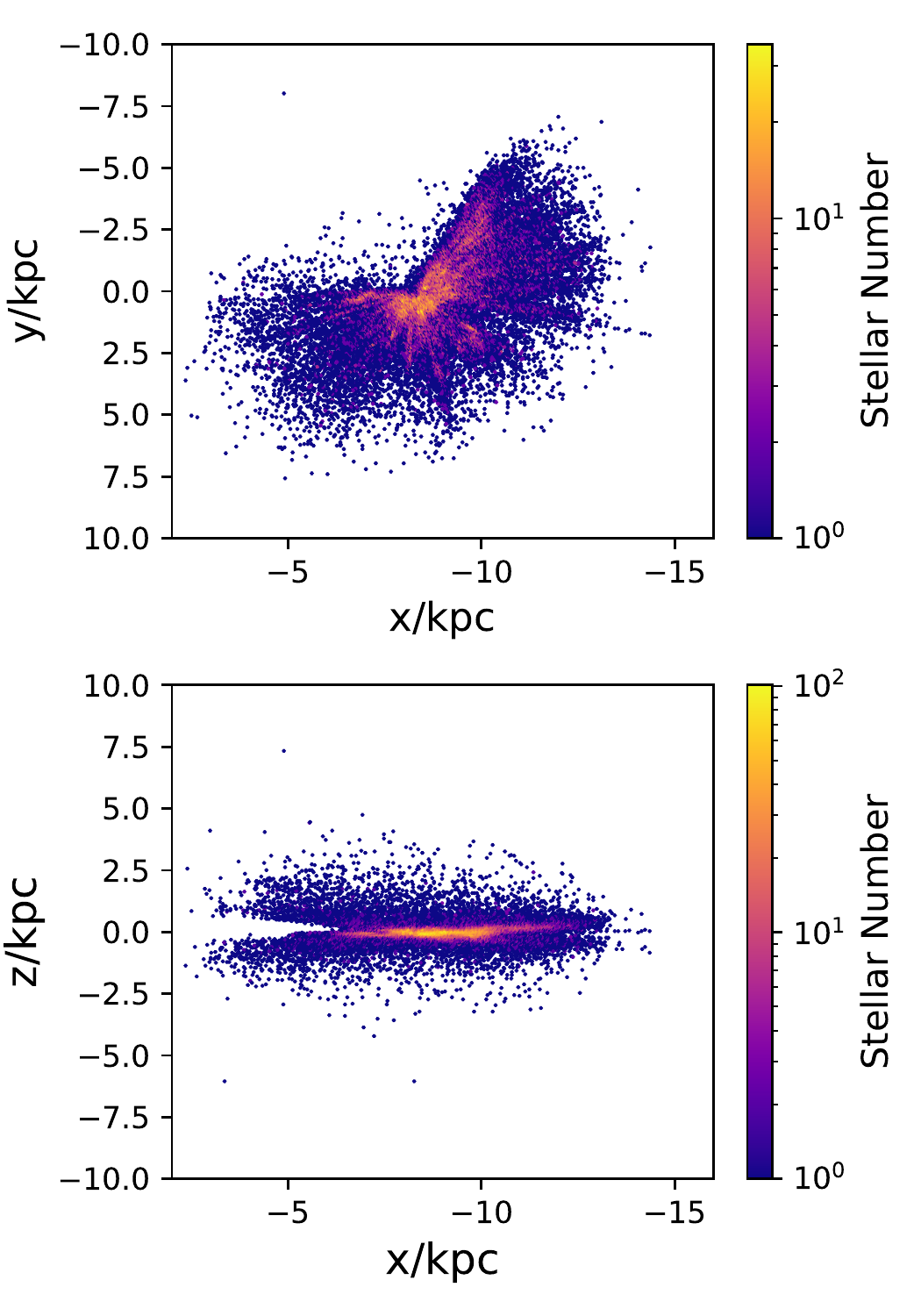}
    \caption{The spatial distribution of our sample of OB stars projected
    onto the  $xy$ plane (upper panel) and $xz$ plane (lower panel).}
    \label{fig:real}
\end{figure}

Finally, we exclude  stars
lying within $3\kpc$ of the Galactic centre to avoid
the barred region of the Galaxy. However, on account of the high extinction
towards the centre and the density profile of the young disc, few stars
are picked even in the range  $R\in(3,5)\kpc$. The final catalogue of OB
stars comprises $46\,916$ stars. They all have apparent magnitudes brighter
than $G=16.65$.
Fig.~\ref{fig:real} shows their  spatial distribution.

\subsection{Selecting OB stars from a model}{\label{sec:selectionfunction}}

\begin{figure}
	\includegraphics[width=\columnwidth]{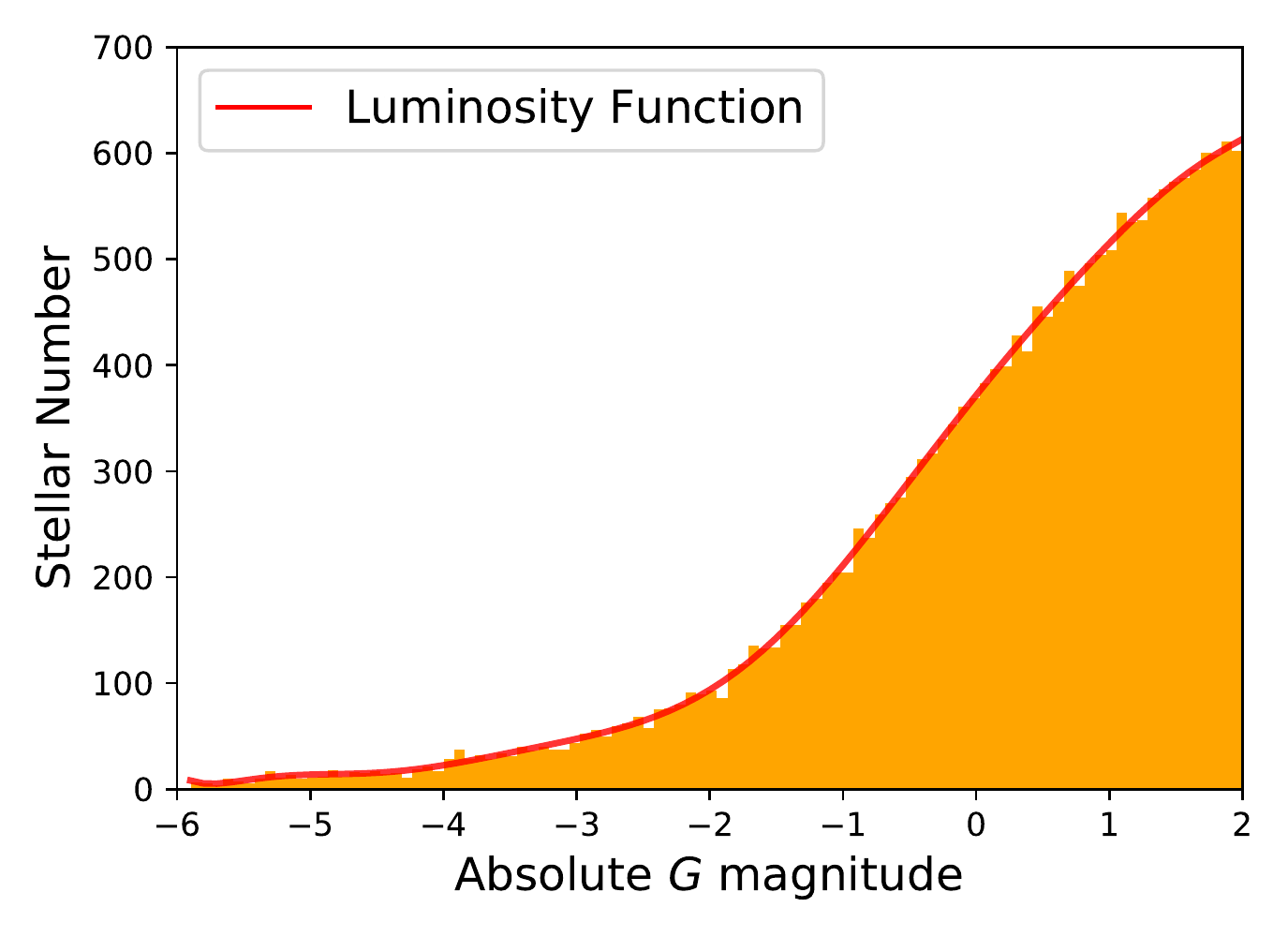}
    \caption{The $G$-band luminosity function for OB stars extracted from the
    PARSEC stellar evolutionary tracks.}
    \label{fig:OBLF}
\end{figure}

Up to an overall normalisation, the DF of the OB stars will differ negligibly
from the DF of the young disc. So the sampling algorithm built into \AGAMA\
can be used to pick a sample of OB stars distributed throughout the Galaxy.
Absolute magnitudes are assigned to these stars by sampling the luminosity
function of OB stars, and using their locations $\vx$ and a dust model, we
compute their apparent magnitudes $G$. The red curve in Fig.~\ref{fig:OBLF}
shows the $G$-band luminosity function that we have used. It was obtained from the
PARSEC stellar evolutionary tracks
\citep{Bressan2012}\footnote{http://stev.oapd.inaf.it/cmd}, which yield the
number of stars expected in a given range of absolute magnitudes per unit
mass of a population with a given star-formation history. We assumed a
constant star-formation rate over the last Gyr. Using this procedure we
obtained luminosity functions for the $\Grp$, $\Gbp$, $J$, $H$ and $K_s$
bands in addition to the $G$ band.

We assign each star observational
uncertainties based on its apparent magnitude\footnote{In fact, this is not
true especially for the proper motions because of the different celestial
frames of Gaia bright and faint stars respectively \citep{CantatGaudin2021}.
However, since the Astrometry Spread Function module in Gaia-verse package
\citep{Everall2021} only works for Gaia DR2 now, the simulated astrometric
solutions are deviated from those in Gaia EDR3. As a result, we use some
simple randomised errors based on apparent magnitudes instead.} and scatter
its phase-space coordinates by these errors. Then a mock star enters the
catalogue if (cf eqn.~\ref{eq:sfgaia})
\begin{align}
\varpi/\epsilon_{\varpi}&>5\cr
M_G&<2\cr
G&<16.65\cr
R&\geq3\kpc,\delta>-30^{\circ}
\end{align}
These criteria are simple because  EDR3 contains essentially all stars
brighter than $G=16.65$. The restriction on $\delta$ arises because we use a
dust model \citep{Green2018,Green2019}. 
that was developed from photometry taken in Hawii \citep{Kaiser2010}.

\section{Formalism}\label{sec:methodology}

Our approach to model fitting is that developed in
\citet{McMillan2012,McMillan2013} and implemented in the case of
five-dimensional data by \citet{Li2022}.  It is based on an algorithm for
determining the likelihood of data given a gravitational potential
$\Phi(\vx)$ a DF $f(\vJ)$ and a selection function $S(\vx,M)$ that gives the
probability that a star with absolute magnitude $M$ located at $\vx$ will
enter the catalogue that is being modelled. The likelihoods are used to drive
a Markov-chain Monte Carlo (MCMC) exploration of the parameter space of DFs.
Two aspects of the present problem require additions to the formulae in
\cite{Li2022}: (i) whereas RR-Lyrae stars can be assumed to have a unique
absolute magnitude, the absolute magnitudes of our OB stars span a
non-negligible range $(M^-,M^+)$, and (ii) each star has an extinction $A$
derived from its location $\vx$. Appendix \ref{app:LiB} derives the additional
formulae.

\section{Distribution function}\label{sec:DF}

BV2022 introduced a new family of DFs for
disc components.  
\begin{equation}\label{eq:discdf}
    f(\vJ)=f_\phi(J_\phi)\,
    f_r(J_\phi,J_r)\,
    f_z(J_\phi,J_z),
\end{equation}
where the functions $f_r$ and $f_z$ are 
\begin{equation}\label{eq:frfz}
    f_i(J_\phi,J_i)=\bigg(\frac{J_{\phi0}}{J_{\rm v}}\bigg)^{p_i}
    \frac{1}{J_{i0}}\,\exp\bigg[-\bigg(\frac{J_{\phi0}}{J_{\rm v}}\bigg)^{p_{i}}\frac{J_i}{J_{i0}}
    \bigg]~~~~(i=r,z).
\end{equation}
This definition involves a function 
\[\label{eq:def_Jv}
J_{\rm v}\equiv |J_\phi|+J_{\rm v0}
\]
 of $J_\phi$ and six  parameters $p_r,p_z,J_{r0},J_{z0},J_{\phi0}, J_{\rm v0}$.
The parameter $p_r$ determines how rapidly the radial velocity dispersion $\sigma_R$
declines outwards, while $p_z$ determines how rapidly $\sigma_z$ declines
with distance from te plane. The constants $J_{r0}$ and $J_{z0}$ set,
respectively, the radial and vertical velocity dispersions. Changes to the constant
$J_{\rm v0}$ modify the central velocity dispersion of the component, which
is not of interest for the present study. 

The form proposed by BV2022 for $f_\phi(J_\phi)$ yields a stellar surface
density that declines monotonically with radius. Since OB stars have formed
recently from molecular gas, and the surface density of gas declines steeply
inside the giant molecular ring, i.e., inside $R\sim5\kpc$, we multiply the
BV2022 form 
\[\label{eq:fphi}
f_\phi(J_\phi)=\frac{M}{(2\pi)^{3}}\,\frac{J_{\rm d}}{2J_{\phi0}^2}\,
    \exp{\bigg(-\frac{J_{\rm d}}{J_{\phi0}}\bigg)}
\]
of $f_\phi$ by
\[\label{eq:taper}
1+\tanh\bigg(\frac{J_{\phi}-|J_{\rm taper}|}{J_{\rm trans}}\bigg),
\]
where $J_{\rm taper}$ is a parameter that determines the radius at which the
surface density peaks and $J_{\rm trans}$ determines the steepness of the
surface density's decline interior to the peak. In the definition
\eqrf{eq:fphi} of $f_\phi$, $M$ is the disc's mass, $J_{\phi0}$ sets the
disc's asymptotic scale-length and 
\[
J_{\rm d}(J_\phi)=|J_\phi|+J_{\rm d0},
\]
where $J_{\rm d0}$ is a constant. Modifying $J_{\rm d}$ changes the disc's
central density, which is not of interest here.

We refer readers to BV2022 for more information on the physical significance
of the DF's parameters. Those that are of concern here are
$p_r,p_z,J_{r0},J_{z0},J_{\phi0}, J_{\rm taper}$ and $J_{\rm trans}$. We fix
$J_{\rm v0}=10\kpc\kms$ and $J_{\rm d0}=20\kpc\kms$.

\section{Potential and dust  models}\label{sec:Gmodel}

Given a DF of the form $f(\vJ)$,  any observable can be predicted once
the Galaxy's
gravitational potential and its  distribution of absorbing dust are
specified.

\subsection{The gravitational potential}\label{sec:potential}

\begin{figure}
	\includegraphics[width=\columnwidth]{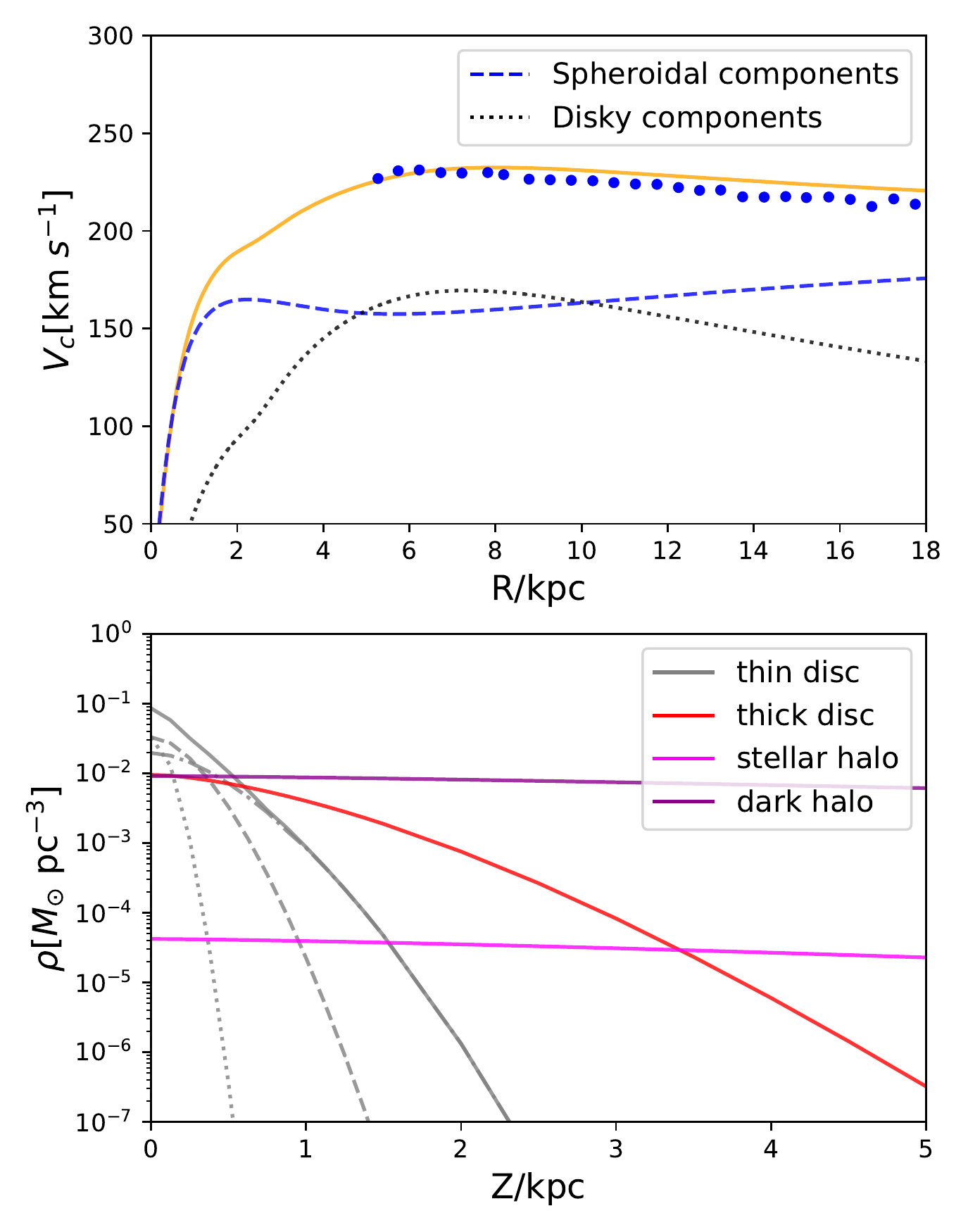}
    \caption{The upper panel shows the circular speed $v_{\rm c}$ of the potential we use. 
The blue dots are observational estimates obtained by \citet{eilers2019}
from red giant stars. The lower panel shows the vertical density
distributions at the solar radius for the components.}
    \label{fig:rotationcurve}
\end{figure}

The upper panel in Fig.~\ref{fig:rotationcurve} shows the circular speed
of the gravitational potential that we have used. The blue dashed line
represents all the spheroidal components including dark halo, bulge, and
stellar halo. The black dotted line denotes the disc components. This
potential was obtained by relaxing to self-consistency an axisymmetric model
Galaxy that is defined by distribution functions for dark matter, a bulge, a
stellar halo, four superposed stellar discs, and a gas disc.
Table~\ref{tab:diskmodel} lists key characteristics of this model.
Further information for the parameters generating this model can be found
in the online supplementary material. The lower panel in
Fig.~\ref{fig:rotationcurve} shows the vertical density distributions of the
components at the solar radius. The dot, dashed, and dash-dotted grey lines
represent young, middle-age, and old components of the thin disc, respectively. The solid grey
line shows the complete thin disc. The red and magenta lines denote thick disc
and stellar halo, respectively. The dark magenta line shows the vertical
density of the dark halo.

\begin{table}
	\centering
	\caption{The disc mass and densities at solar radius.}
	\label{tab:diskmodel}
	\begin{tabular}{lccc} 
		\hline
		$M_{\rm disk}$       & $5.48\times10^{10}\mbox{M}_{\odot}$ \\
		$\rho(R_0,z=0)$& 0.095 $\mbox{M}_{\odot}\mbox{pc}^{-3}$    \\ 
		$\rho(R_0,z=1)$& 0.005 $\mbox{M}_{\odot}\mbox{pc}^{-3}$   \\
		\hline
	\end{tabular}
\end{table}

\subsection{Observational dust model}\label{sec:dust}

\begin{figure}
	\includegraphics[width=\columnwidth]{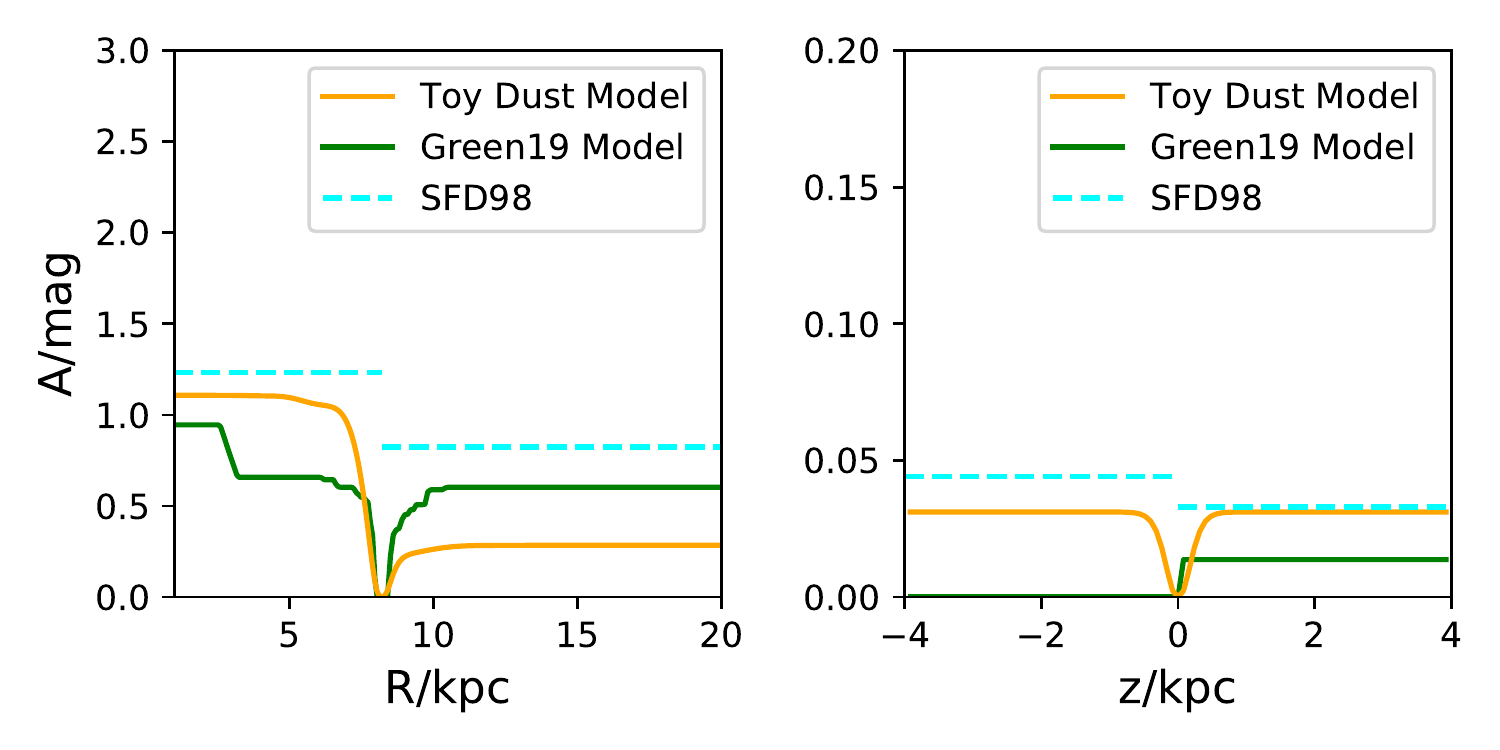}
    \caption{estimates of extinction versus distance (i) in the centre and
    anti-centre directions at $b=10\,$deg (left panel)  and verticallu up or
    down (right panel). The green and orange lines are from the \textit{Bayestar2019}
\citep{Green2019} and toy models, respectively.  The dotted cyan lines show
the asymptotic values  from \citet{Schlegel1998}.}
    \label{fig:ext_compare_y=0}
\end{figure}

\begin{figure}
	\includegraphics[width=\columnwidth]{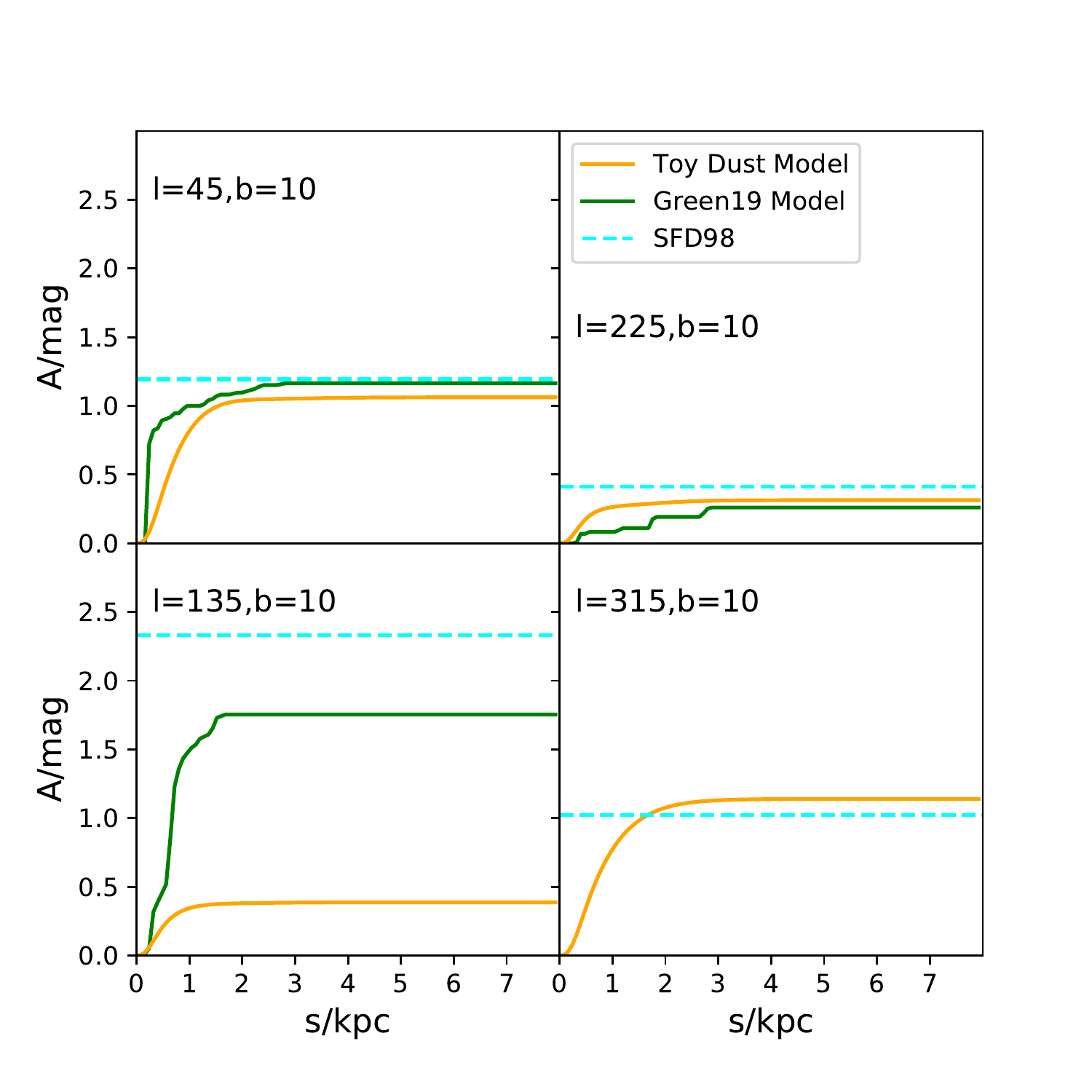}
    \caption{As Fig.~\ref{fig:ext_compare_y=0} for lines of sight at
    $b=10\,$deg and various longitudes.}
    \label{fig:ext_compare_4}
\end{figure}

\begin{figure*}
	\includegraphics[scale=0.45]{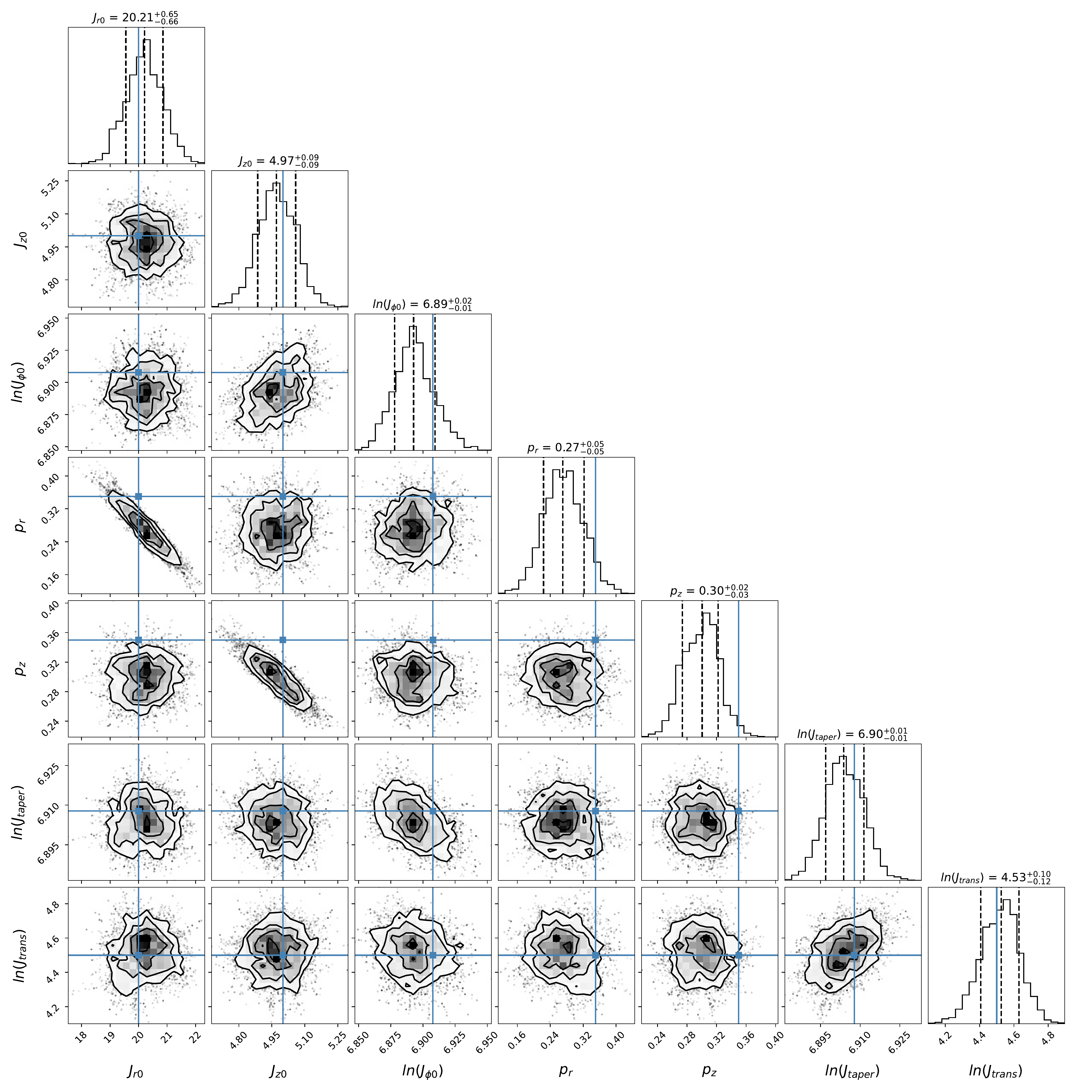}
    \caption{Posterior probability distributions from a  mock catalogue.}
    \label{fig:mock_fit_5kpc}
\end{figure*}

Several groups have recently developed models of the Galactic distribution of
dust \citep{Sale2014,Lenz2017,Green2018,Green2019}. The green curves in
Figs.~\ref{fig:ext_compare_y=0} and \ref{fig:ext_compare_4} show extinction
as a function of distance along eight lines of sight according to the
\textit{Bayestar2019} model from the most recent of these studies. The left
panel of Fig.~\ref{fig:ext_compare_y=0} shows the line of sight
$(\ell=0,b=10)$ towards the Galactic centre and that $(\ell=180,b=10)$ towards
the anti-centre, while the right panel shows the lines of sight vertically
downwards and upwards. The upper panels of Fig.~\ref{fig:ext_compare_4} are
for lines of sight at $45\,$deg to the centre- and anti-centre directions
while the bottom left panel is for the direction that in an axisymmetric
Galaxy would be equivalent to that of the middle right panel.  Actually, the
asymptotic extinction is about six times larger at $\ell=135\,$deg than at
$\ell=225\,$deg.  In each panel the cyan line shows the extinction towards
extra-Galactic objects from \cite{Schlegel1998}.  Ideally, each green curve
would asymptote at large distances to its cyan line.  Unfortunately the
reality falls well short of this ideal with the value from \cite{Green2019}
systematically smaller than that from \cite{Schlegel1998}. 

\subsection{Toy dust model}\label{sec:toyDust}

Since Figs.~\ref{fig:ext_compare_y=0} and \ref{fig:ext_compare_4} suggest
that even the best current extinction map is likely far from the truth, we
have investigated how results obtained from mock data are affected by use of
an incorrect dust model. For these tests we used a toy dust distribution
assembled by adding a spiral distribution and a local bubble to an underlying
axisymmetric distribution of dust.  The latter is
\begin{equation}\label{eq:toy_dust}
    \rho(R)=\rho_0\,\exp{\Big(-\frac{R}{R_s}-\frac{z}{z_s}}\Big),
\end{equation}
where $R_s=3\kpc$ and $z_s=0.1\kpc$ are scale length and
height of the dust disc, and $\rho_0=5$ is a
scale density. We add a four-arm spiral pattern to  this  axisymmetric
density distribution by multiplying it by
\begin{equation}\label{eq:toy_spiral}
    s=1+\beta \cos^{2}{(m h_{s}/2)},
\end{equation}
where $\beta=6, m=4$ and
\begin{equation}\label{eq:hs}
    h_s=\alpha \ln{R}+(\phi-\phi_0),
\end{equation}
with $\alpha=4,\phi_0=135^{\circ}$, and $R=\sqrt{x^2+y^2}$. 

To simulate the local bubble we further multiply the above density by
\begin{equation}\label{eq:localbubble}
    b=1-\exp{(-s^2/2\sigma^2)},
\end{equation}
where $s$ is heliocentric distance and $\sigma=0.2\kpc$ sets the scale  of the bubble.

The extinction is then the integral  $A=\int\d s\,\rho(R,z,\phi)$ of
$\rho(R,z,\phi)$ along the line of sight and we use this extinction to assign
apparent magnitudes to mock stars proposed by \AGAMA. When analysing the
resulting
mock catalogue, we use the dust model from \cite{Green2019}, which is
in this case inaccurate.

The orange curves in Figs.~\ref{fig:ext_compare_y=0} and
\ref{fig:ext_compare_4} show an example of extinctions from this model
alongside the empirical extinctions from \citet{Green2019} and
\citet{Schlegel1998}. The most serious defect of the toy model is a failure
to generate radically different extinctions at the longitudes
$\ell=180\pm45\,$deg. This failure is attributable to its spherical local
bubble rather than a cavity that is bounded by a very lumpy wall
\citep{Zucker2022}. However, even a simplistic toy model of the dust enables
us to probe the consequences of fitting data using a defective dust model.

\section{Tests}\label{sec:tests}

\begin{table*}
	\centering
	\caption{Summary of test results.
	Actions are given in $\kpc\kms$.}
	\label{tab:tests}
	\begin{tabular}{lccccccc} 
		\hline
		Parameters & $J_{r0}$ & $J_{z0}$ & $\ln{J_{\phi0}}$ &
		$p_{r}$ & $p_{z}$ & $\ln{J_{{\rm taper}}}$ & $\ln{J_{{\rm trans}}}$\\
		\hline
		Input Values & 20 & 5 & 6.91 & 0.35 & 0.35 & 6.91& 4.50\\
		Single catalogue & $20.21^{+0.65}_{-0.66}$ & $4.97^{+0.09}_{-0.09}$ & $6.89^{+0.02}_{-0.01}$ & $0.27^{+0.05}_{-0.05}$ & $0.30^{+0.02}_{-0.03}$ & $6.90^{+0.01}_{-0.01}$& $4.53^{+0.10}_{-0.12}$\\
		Mean of 10 catalogues& $19.07\pm0.88$ & $5.01\pm0.10$ & $6.91\pm0.01$ & $0.31\pm0.06$ & $0.33\pm0.03$ & $6.90\pm0.01$ & $4.42\pm0.09$\\
		Wrong dust model& $16.87^{+1.15}_{-1.82}$ & $4.18^{+0.13}_{-0.12}$ & $6.76^{+0.03}_{-0.03}$ & $0.48^{+0.14}_{-0.05}$ & $0.39^{+0.04}_{-0.04}$ & $7.01^{+0.02}_{-0.02}$& $5.11^{+0.12}_{-0.19}$\\
		\hline
	\end{tabular}
\end{table*}

We generated ten mock catalogues as described in Section
\ref{sec:selectionfunction}. The parameters used to generate these catalogues
are given in the top row of Table~\ref{tab:tests}. Then  the computer
explored the likelihood in the seven-dimensional parameter space with
coordinates $(J_{r0},J_{z0},\ln J_{\phi0},p_r,p_z,\ln J_{\rm tapper},\ln
J_{\rm trans})$. Thus we were holding fixed the parameters $J_{\rm d0}$ and
$J_{\rm v0}$ that only affect a model in the region interior to all the data.
In some runs these two parameters took the values used to generate the mock data,
while in other runs they were fixed at values that differed from those that
generated the data. These tests showed insensitivity of results to
the values of these two parameters.

The \textit{SLSQP} method in the \textbf{scipy} \citep{Virtanen2020} package
was first used to find the maximum of the likelihood and then the
$\textit{emcee}$ package \citep{Mackey2013} was used to run a Markov Chain
Monte-Carlo exploration starting from the maximum likelihood.  We used 35
walkers for each parameter over a total of 500 steps with the first 100 steps
treated as burn-in.\footnote{The convergence was tested by computing the
auto-correlation time recommended by \textit{emcee}. The average
auto-correlation time for all the parameters are about 25 which means the
burn-in steps are reasonable. We also test a longer chain with $N=3000$ which
is more than 100 times the auto-correlation time. It yielded similar results
to those obtained with $N=500$ and in the interests of economy we $N=500$
here.} Fig.~\ref{fig:mock_fit_5kpc} shows a typical set of projections of the
posterior probability distribution.  The blue lines in each panel indicate
the true parameter values while the dashed lines in the histograms show the
1-$\sigma$ uncertainties from the 16th and 84th percentiles. The second row of
Table~\ref{tab:tests} lists these uncertainties of the posterior distribution
yielded by one catalogue, while the table's third row gives the means and
standard deviations of the means of the ten posterior distributions. The
standard deviations are close to the 1-$\sigma$ uncertainties from individual
posterior distributions, but in most cases are slightly larger, as would be
expected if the distributions had longer tails than a Gaussian.

Only $p_r$ and $p_z$ have true values that lie outside the 1-$\sigma$
range inferred from  a single catalogue and even these true values lie within
the 1-$\sigma$ ranges inferred from ten catalogues. Thus the MCMC runs are
delivering consistent results. Moreover, the precision with which parameters
can be recovered is impressive -- the uncertainty of the actions other than
$J_{r0}$ is $\la2$
per cent, while that of $J_{r0}$ is better than 4 percent. 

\subsection{Correlations between parameters}{\label{sec:mock_parameter}}

\begin{figure}
	\includegraphics[width=\columnwidth]{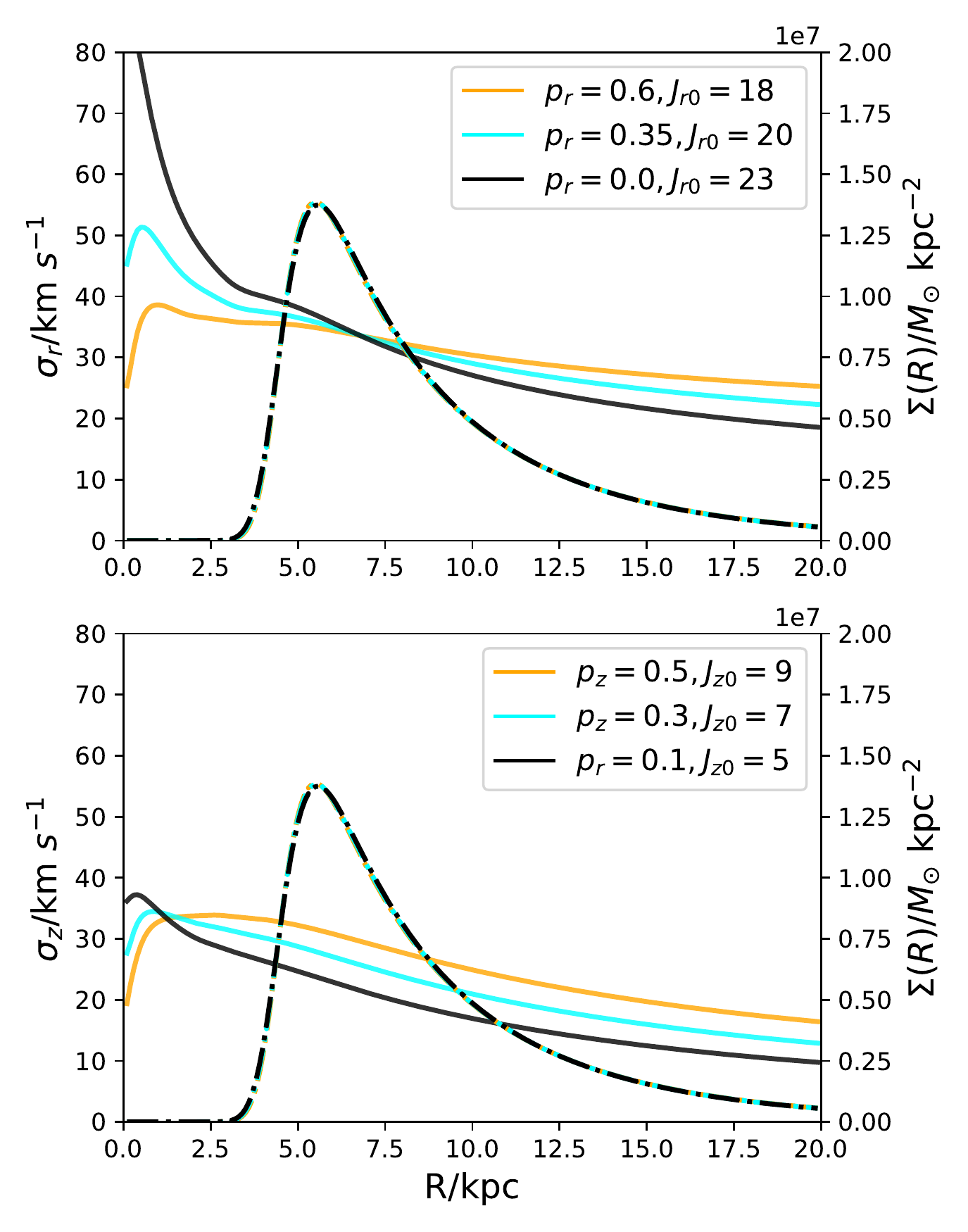}
    \caption{Full lines: the radial and vertical velocity dispersions $\sigma_R$ and
    $\sigma_z$ as functions of radius $R$ in the Galactic plane for several
models. Broken lines: the surface density $\Sigma(R)$ of OB stars -- the
densities of all models considered are indistinguishable.}
    \label{fig:vsigma2}
\end{figure}

Fig.~\ref{fig:mock_fit_5kpc} shows just two significant correlations between
parameters, namely between $J_{r0}$ and $p_r$ and between $J_{z0}$ and $p_z$.
$J_{r0}$ sets the magnitude of the radial velocity dispersion $\sigma_R$,
while $p_r$ sets the radial gradient of $\sigma_R$. $J_{z0}$ and $p_z$
similarly set
the magnitude and radial gradient of $\sigma_z$.  In the upper panel of
Fig.~\ref{fig:vsigma2} we plot $\sigma_R$ versus $R$ for three DFs. Also
plotted as a chained black line is the radial density of mock stars in the
catalogue.  We see that very similar values of $\sigma_R$
are obtained within the well sampled radial range using values of $p_r$ that
range from 0 up to $0.6$ by compensating for the change in $p_r$ by a change
in $J_{r0}$. Interior to the well sampled range the three models predict very
different values for $\sigma_R$, so they could be distinguished if the catalogue
contained a significant number of stars at $R<3\kpc$. The lower panel of
Fig.~\ref{fig:vsigma2} makes the same point in the context of $\sigma_z$.
Hence the strong degeneracies evident in Fig.~\ref{fig:mock_fit_5kpc} reflect
the lack of stars at $R\la3\kpc$.

The uncertainties in $J_{z0}$ and $p_z$ are smaller than those in $J_{r0}$ and
$p_r$ because the greatest differences in $R$ are achieved towards the
anti-centre, where $v_R$ dominates $v_\parallel$, which is not in the
catalogue, while $v_z$ dominates $v_b$, which is in the catalogue.

\subsection{Impact of the dust model}{\label{sec:dust_model}}

Now we explore how using an incorrect dust model affects the posterior
distribution  of  $J_{\phi0}$ and $J_{\rm taper}$. 
Fig.~\ref{fig:mock_wrong_dust} shows the posterior probabilities obtained
when a mock catalogue created using the toy dust model of
Section~\ref{sec:toyDust} is analysed using the dust model of
\cite{Green2019}. The bottom row of Table~\ref{tab:tests} lists
the corresponding statistics.

The formal uncertainties on the parameters, $\ln J_{\phi0}$, $\ln J_{\rm
taper}$ and $\ln J_{\rm trans}$ that determine the stellar surface density
increase only moderately but the true value of $\ln J_{\phi0}$ now lies
$0.12$ above its 1-$\sigma$ range, while the true values of $\ln J_{\rm
taper}$ and $\ln J_{\rm trans}$ now lie $0.08$ and $0.42$ below their
1-$\sigma$ ranges.  The lower left panel of Fig.~\ref{fig:ext_compare_4} shows that around
$\ell=135\,$deg, the toy model predicts much lower
extinction than the \cite{Green2019} model. Low extinction enhances the number of mock
stars, and to account for these stars in the presence of
the larger observed extinction, the disc must be imagined more extended than
it is. Hence, the recovered value of $J_{\phi0}$ should be larger than the true
value. That this is not the case must be attributed to the unrealistically
large values of $J_{\rm taper}$ and $J_{\rm trans}$: increasing $J_{\rm
taper}$ pushes outwards the flat part of the surface density, and increasing
$J_{\rm trans}$ extends the flat section.

The formal uncertainties on $J_{r0}$ ,$J_{z0}$, $p_r$ and $p_z$ all increase
significantly and the true values of both $J_{r0}$ and $J_{z0}$ now lie
above their 1-$\sigma$
ranges, while the true values of $p_r$ and $p_z$ still lie within their
1-$\sigma$ ranges. 

This experiment strongly suggests (i) that the formal uncertainties on the
recovered parameters of the DF materially under-estimate the true
uncertainties because the latter are dominated by the uncertain distribution
of dust, and (ii) that degeneracies between the parameters make it hard to
predict how recovered parameters will be affected by a change in the dust
model.

\section{Fits to Gaia data}\label{sec:fits}

\begin{figure*}
	\includegraphics[scale=0.45]{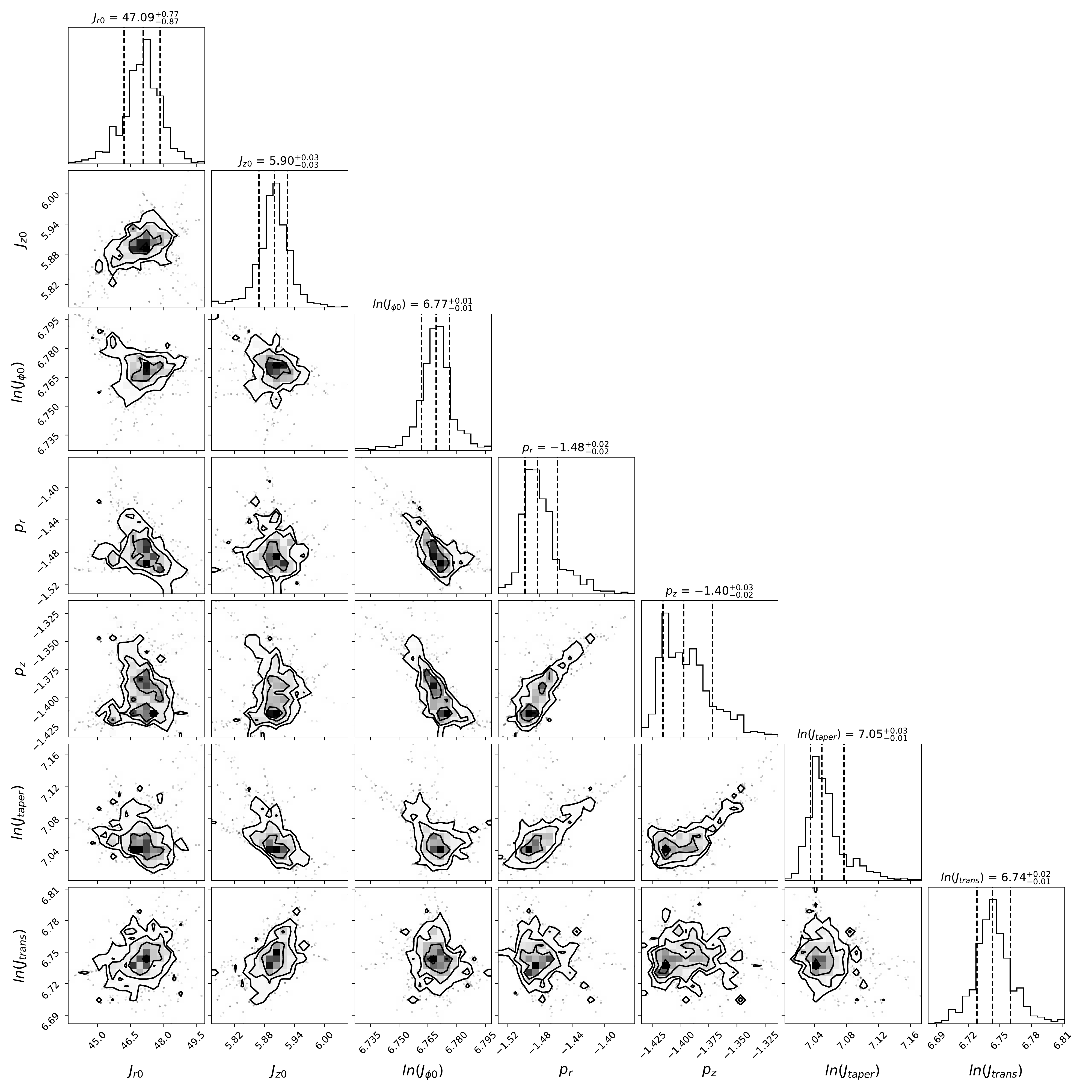}
    \caption{Posterior probability distributions from the real catalogue.}
    \label{fig:real_result}
\end{figure*}

\begin{table*}
	\centering
	\caption{Statistics of the probability distributions from fits to
	Gaia OB stars.  $J_{d0}$ and $J_{v0}$ are kept fixed at $20\kms$ and
	$10\kms$, respectively.}
	\label{tab:real_result}
	\begin{tabular}{lccccccccc}
		\hline
		Parameters & $J_{r0}$ & $J_{z0}$ & $\ln{J_{\phi0}}$ & $p_{r}$
		& $p_{z}$ & $\ln{J_{\mbox{taper}}}$ & $\ln{J_{\mbox{trans}}}$& $J_{d0}$ & $J_{v0}$\\
		\hline
		Fitted result& $47.09^{+0.77}_{-0.87}$ & $5.90^{+0.03}_{-0.03}$ & $6.77^{+0.01}_{-0.01}$ & $-1.48^{+0.02}_{-0.02}$ & $-1.40^{+0.03}_{-0.02}$ & $7.05^{+0.03}_{-0.01}$& $6.74^{+0.02}_{-0.01}$& 20 & 10 \\
		\hline
	\end{tabular}
\end{table*}

When the code was used for an MCMC search of fits to the real data, the
chain showed a long-term drift towards larger $J_{r0}$ and smaller
$J_{\phi0}$. In light of this drift, the MCMC chain was extended to 1500
steps, three times the number of steps used for the mock data, but without
convincing evidence emerging that the chain's long-term drift had ceased. As
discussed below, we think this poor performance of the code on real data
is a
consequence of our use of an erroneous dust model. On account of the latter,
no DF gives a convincing fit to all the data. Different DFs fit
different parts of the data and the code wanders from one unsatisfactory
DF to another without finding a convincing maximum in the data's
likelihood.

Here we present results from a near-converged section of $\sim 300$ steps.
These results illustrate the quality of fit to the data that can be achieved
with the present dust model, and provide an indication of the parameter
choices that might be found when a satisfactory dust model becomes
available.
Fig.~\ref{fig:real_result} shows posterior distributions from the chosen
section of the chain, while Table~\ref{tab:real_result} gives the means and
1-$\sigma$ uncertainties of these distributions.  It should be borne in mind
that the true uncertainties will be significantly larger than those quoted
here on account of the premature truncation of the MCMC chain.

In Fig.~\ref{fig:real_result}, the histograms are less Gaussian than the
corresponding histograms in Fig.~\ref{fig:mock_fit_5kpc}, and in the
off-diagonal panels the regions of
high likelihood are typically much more elongated than the corresponding
regions. We attribute both phenomena to the failure of the chain to converge
on account of the poor dust model.

\subsection{Fit quality}\label{sec:quality}
\begin{figure}
	\includegraphics[width=\columnwidth]{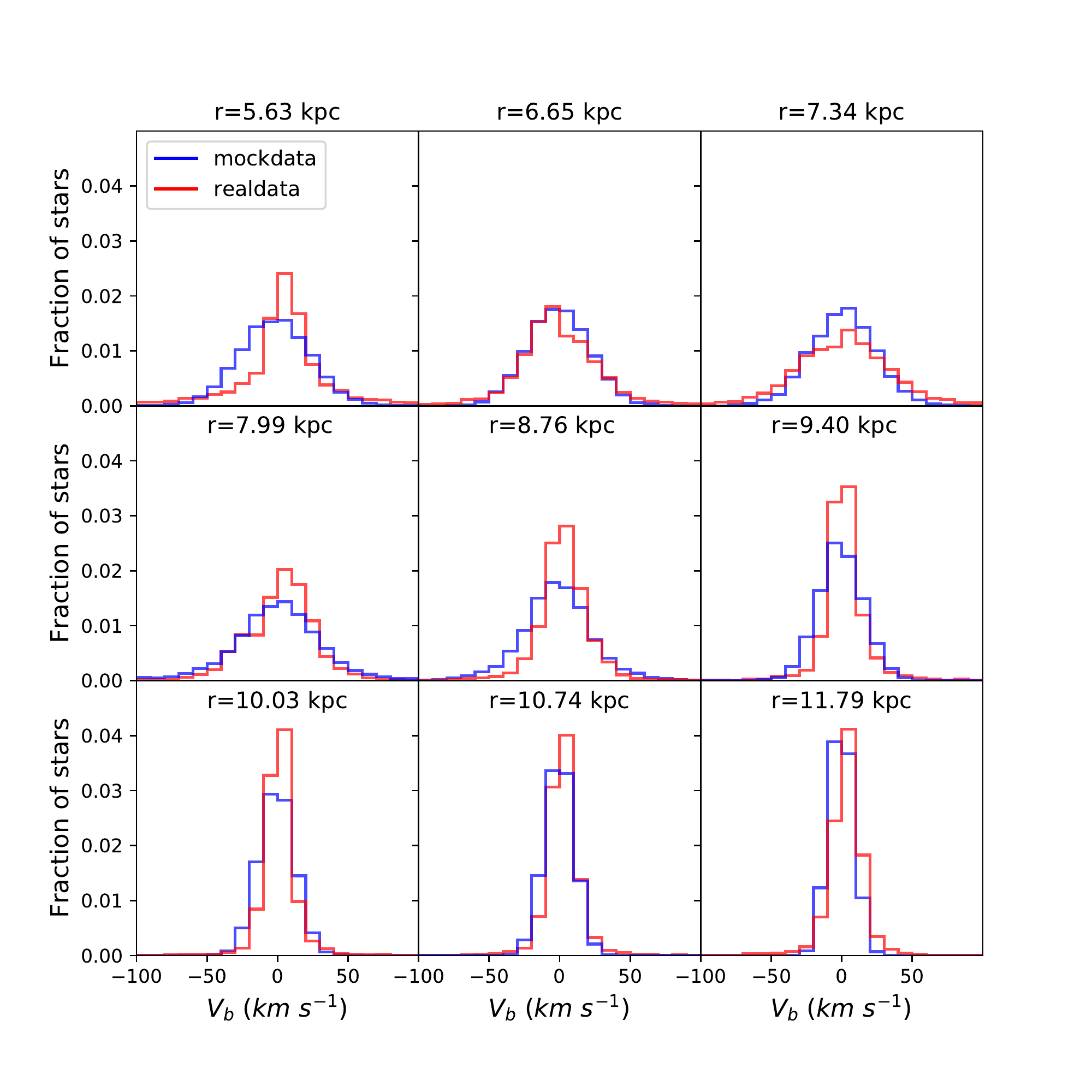}
    \caption{Histograms of $v_b$ for real data (red) and for mock stars drawn
    from
    the best-fit model. For each bin we give  the median $R$ value of the
    mock stars.}
    \label{fig:vb}
\end{figure}

\begin{figure}
	\includegraphics[width=\columnwidth]{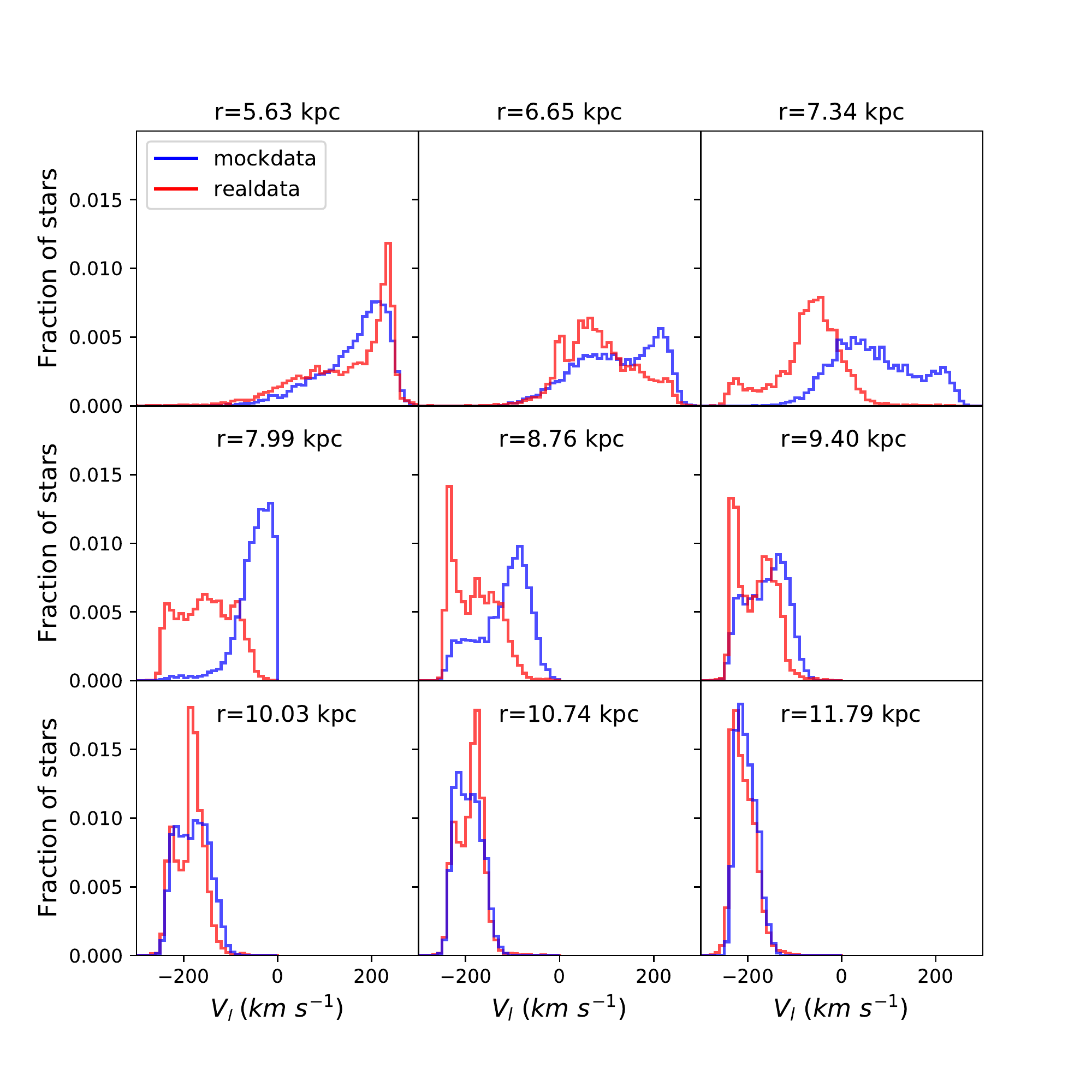}
    \caption{As Fig.~\ref{fig:vb} but for $v_\ell$.}
    \label{fig:vl}
\end{figure}

\begin{figure}
	\includegraphics[width=\columnwidth]{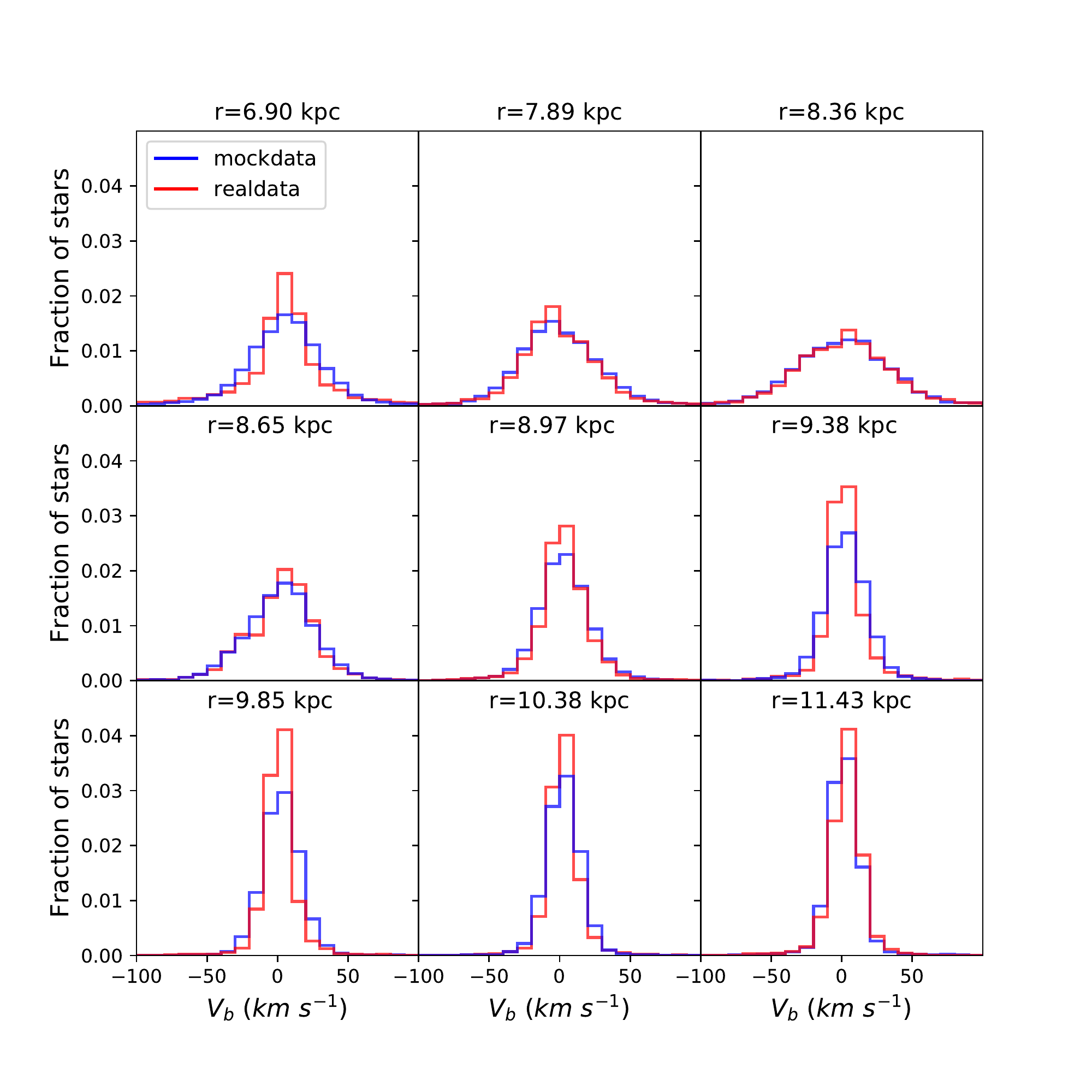}
    \caption{As Fig.~\ref{fig:vb} except that the blue histogram is computed
    from 
    velocities assigned by sampling the best-fit model at the locations of real rather
    than mock stars.}
    \label{fig:vb_sample}
\end{figure}

\begin{figure}
	\includegraphics[width=\columnwidth]{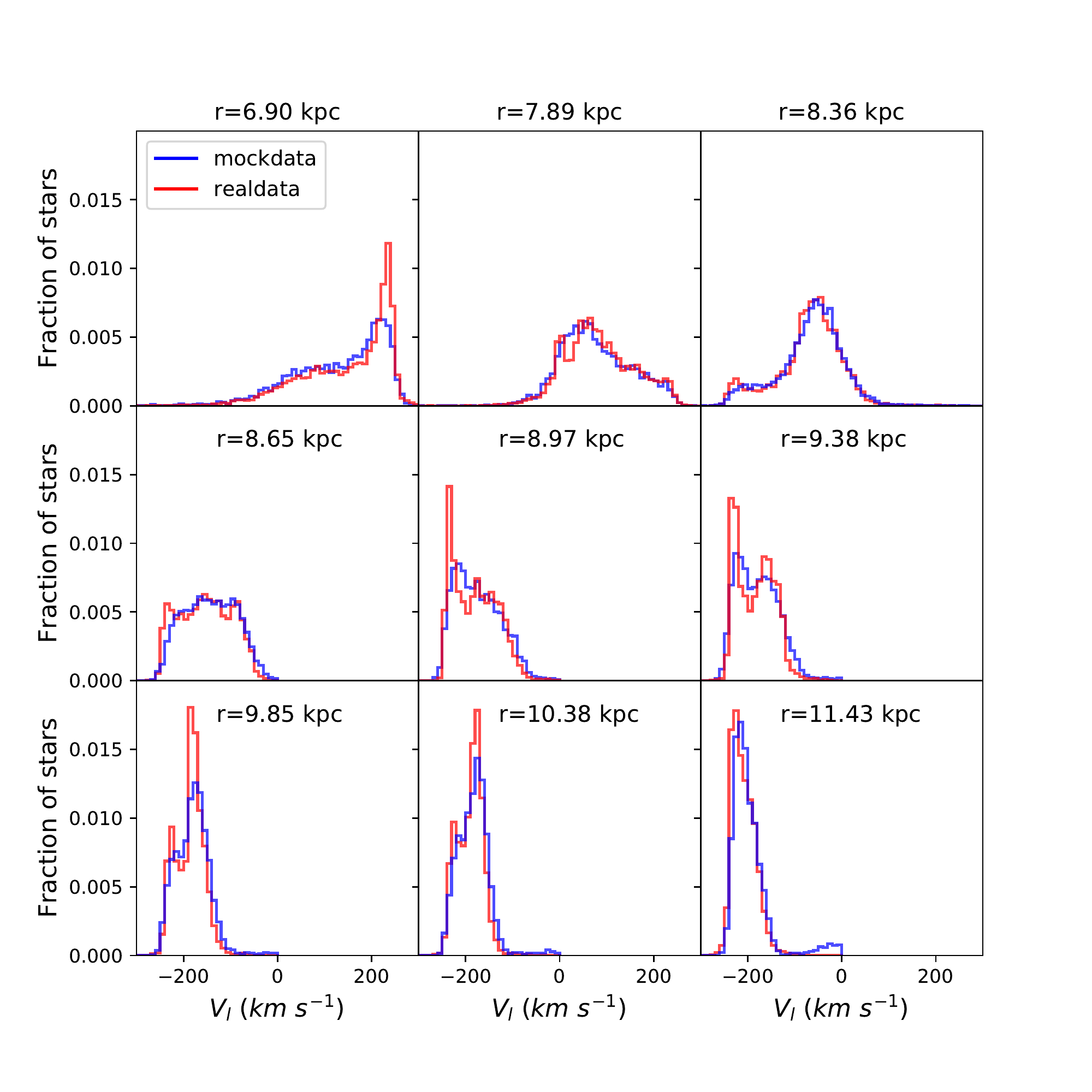}
    \caption{As Fig.~\ref{fig:vb_sample} for the $v_\ell$ component.}
    \label{fig:vl_sample}
\end{figure}

A natural first question is whether the  models with the largest recovered
likelihoods give an acceptable account of the data. Figs.~\ref{fig:vb}
and \ref{fig:vl} are histograms of velocity in the $b$ and $\ell$ directions,
respectively for stars binned by radius. The red histograms are for real
stars and the blue histograms are for a sample of mock stars drawn from the
model that gives the real data the highest likelihood. While the red and blue
histograms agree moderately well in Fig.~\ref{fig:vb} for $v_b$, they are
disturbingly different in Fig.~\ref{fig:vl} for $v_\ell$.
Figs.~\ref{fig:vb_sample} and \ref{fig:vl_sample} shed light on these
differences by showing analogous histograms when we use the same model to
choose a velocity for each real star. Now the real and mock  distributions
agree at least as well in $v_\ell$ as in $v_b$. Evidently, the clash in
Fig.~\ref{fig:vl} arises because the real and mock stars have systematically different
locations; at a given place, the model predicts velocities accurately, but it
makes poor choices for the locations of stars.

\begin{figure}
	\includegraphics[width=\columnwidth]{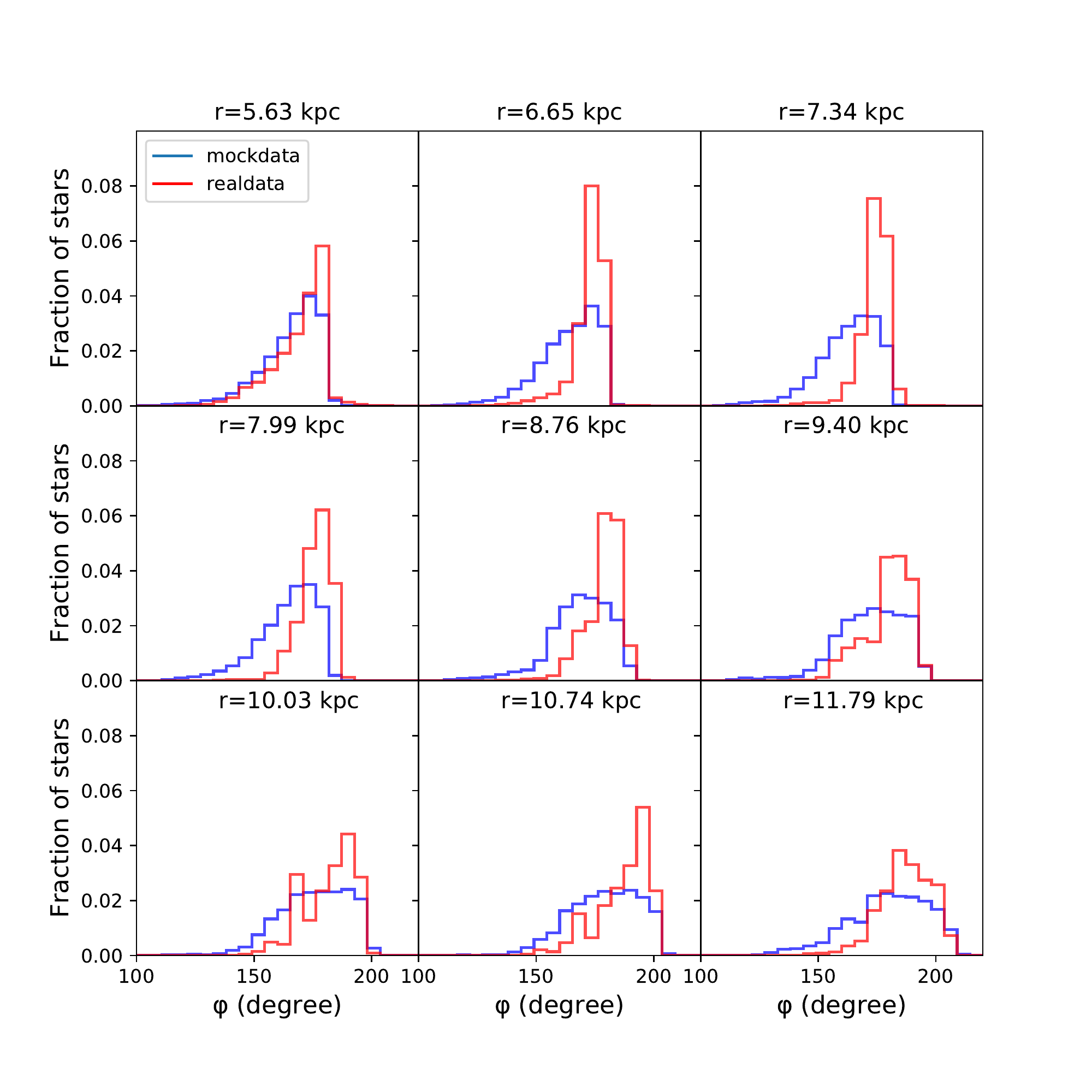}
    \caption{The distribution of $\phi$ for both best-fit mock and real
catalogue stars. The bins are the same as for Figs.~\ref{fig:vl} and
\ref{fig:vb}. }
    \label{fig:phi9}
\end{figure}

Fig.~\ref{fig:phi9} drives this point home by showing the real and mock
distributions of stars in the Galactocentric angle $\phi$. In general the
real distribution is narrower and shifted to larger $\phi$ than the mock
distribution. Since differential rotation of the disc makes $v_\ell$ a strong
function of $\phi$, different distributions in $\phi$ inevitably lead to a
different distributions in $v_\ell$.

The prime suspect for causing these differences in the distribution of $\phi$
between real and mock stars must be the dust model. Indeed, our axisymmetric
DF has very little capacity to modify the distribution in $\phi$.
Figs.~\ref{fig:ext_compare_y=0} and \ref{fig:ext_compare_4} by contrast
demonstrate that dust violently breaks the symmetry in $\phi$.  If the dust
model breaks this symmetry incorrectly, the mock stars will not reproduce the
observed bias in $\phi$, and the predicted values of $v_\ell$ will be wrongly
distributed.

\begin{figure}
	\includegraphics[width=\columnwidth]{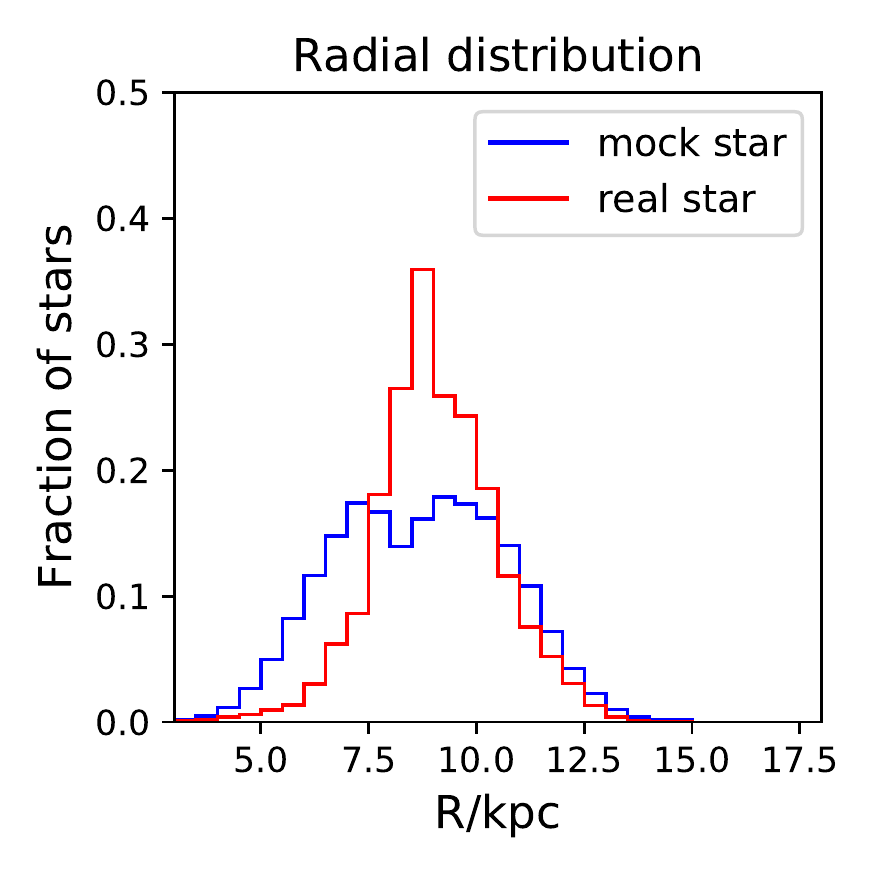}
    \caption{The distribution in $R$ of the real stars (red) and mock stars
    (blue) drawn from the best-fit model.}
    \label{fig:rphi}
\end{figure}

Of course dust also has a strong influence on the radial distribution of
catalogued stars.  The red and blue histograms in Fig.~\ref{fig:rphi} show
the real and mock distributions in $R$. The mock distribution is broader than
the real one because it extends to smaller radii. This finding suggests that
the extinctions provided by the  dust model tend to be too small in
directions towards the Galactic Centre.

Changing the radial distribution of OB stars is very much within the scope of
the DF, so it is perhaps puzzling that the best-fitting model does not
reproduce the observed distribution in $R$ better. The answer may be that
when the DF changes the radial distribution of stars, for example by changing
the value of $J_{\rm taper}$, it will also change the kinematics. So the need
to fit the observed kinematics can push the DF to predict (correctly) an
abundance of stars at small radii that do not feature in the data, but do in
a mock catalogue drawn using extinctions that are too small.

\begin{figure}
	\includegraphics[width=\columnwidth]{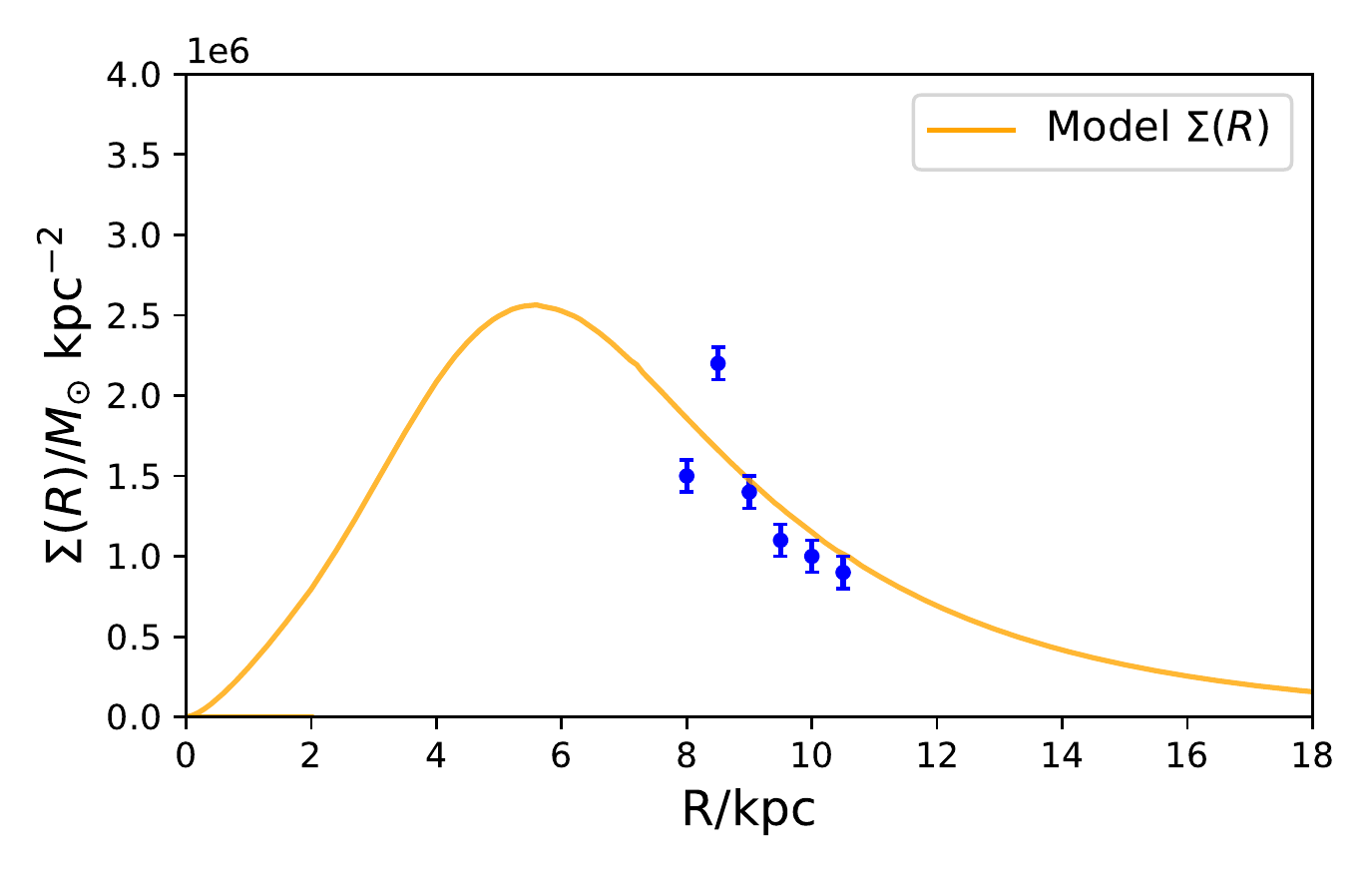}
    \caption{The curve shows the surface density of the best-fit model of
    the young disc. The blue data points show the surface densities derived
    by \citet{Xiang2018} from LAMOST data. The
normalisation of the model is reduced by a factor 3 to facilitate the
comparison of data and model.}
    \label{fig:surface_density}
\end{figure}

The curve Fig.~\ref{fig:surface_density} shows the surface density of the
best-fitting model. The density peaks at $R\sim5.5\kpc$, which is now thought
to be close to the bar's corotation radius
\citep{Perez2017,Binney2020,Chiba2021}. The density drops faster from the peak
inwards than outwards. The blue dots in Fig.~\ref{fig:surface_density} show
the surface density of the young disc estimated by \citet{Xiang2018} using a
catalogue from the LAMOST survey. They found a significant peak at the solar
radius which they suggests reflects the Local Arm but they were unable to
probe the disc interior to the Sun on account of LAMOST's bias towards the
anticentre.  \citet{Bovy2016} used red clump stars in APOGEE to assess the
structure of mono-abundance populations (MAPs). The metal-rich, low-$\alpha$
population to which the OB stars belong, was found to have a ``broken"
exponential radial profile, with the break at $R=7\kpc$.
\citet{Mackereth2017}, using red giants in APOGEE as the tracers, found that
the youngest population has a break radius at $R\simeq8\kpc$. The surface
density of our best-fitting model shown in Fig.~\ref{fig:surface_density} is
broadly in line with these earlier results, although its break radius is
somewhat smaller. 

The predicted peak in the surface density of OB stars at $R\simeq5.5\kpc$
falls outside the reported peak at $R=4-5\kpc$ in the surface density of
H$_2$ \citep{Heyer2015}.  While studies of the stellar disc are liable to
over-estimate the radius of the peak stellar density because the innermost
young disc is hidden behind the abundant dust near the plane at $R\la6\kpc$,
the radius at which the density of H$_2$ peaks may have been under-estimated.
Indeed, hydrodynamical models \citep[e.g.][]{SormaniIII} predict the density
of gas to be low inside the bar's corotation radius, which is currently
believed to be as large as $6\kpc$ \citep{Perez2017,Binney2020,Chiba2021}. A
circular-speed curve is required to convert an observed distribution of
emission-line intensity in longitude and velocity into a plot of surface
density versus radius. Perhaps when the emission-line data are re-analysed
using a circular-speed curve extracted from Gaia, the H$_2$ distribution will
be found to peak at a larger radius.

\subsection{Comparison the LAMOST}\label{sec:LAMOST}

Our model predicts distributions of space velocities and in this section we
compare these predictions with distributions inferred by combining
spectroscopic measurements of $v_\parallel$ with EDR3 astrometry.
Specifically, many of the OB stars we have identified in the intersection of
EDR3, 2MASS and the Starhorse catalogue were observed by LAMOST
\citep{Xiang2021} so have measured values of $v_\parallel$.  We select stars
within $\left| b\right| <5\,$deg and $\epsilon_v<50\kms$ to exclude stars
with poorly determined $v_\parallel$. We also shift the LAMOST values of
$v_\parallel$ by $4.54\kms$ as suggested by
\citet{Schonrich2017,Anguiano2018}.  After such corrections, LAMOST's values
of $v_\parallel$ agree well with the values measured for the same stars by
APOGEE and Galah. 

\begin{figure*}
\includegraphics[scale=0.55]{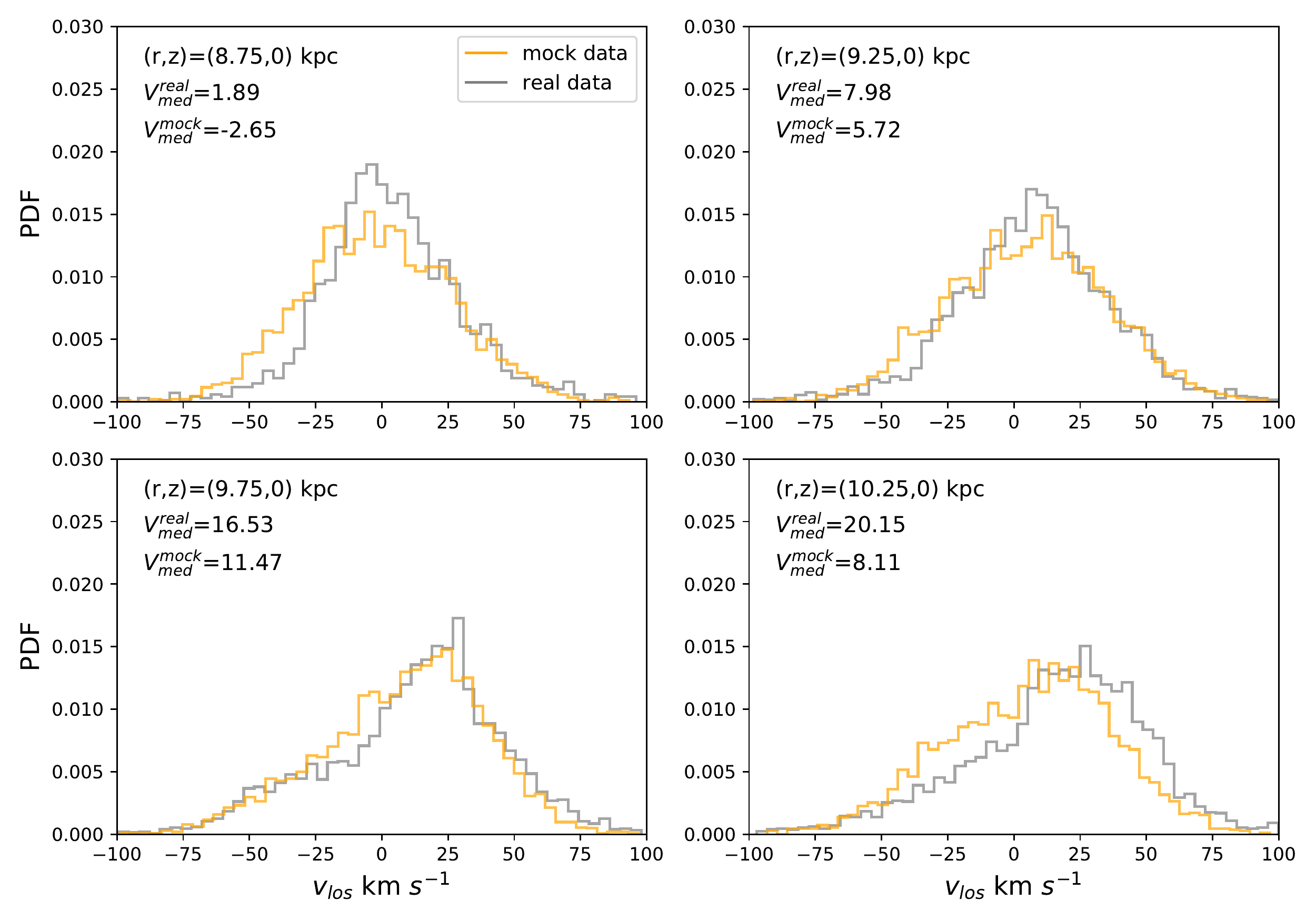}
\caption{Observed and mock histograms of $v_\parallel$ in four bins in
$R$}\label{fig:Vlos}
\end{figure*}

\begin{figure*}
	\includegraphics[scale=0.55]{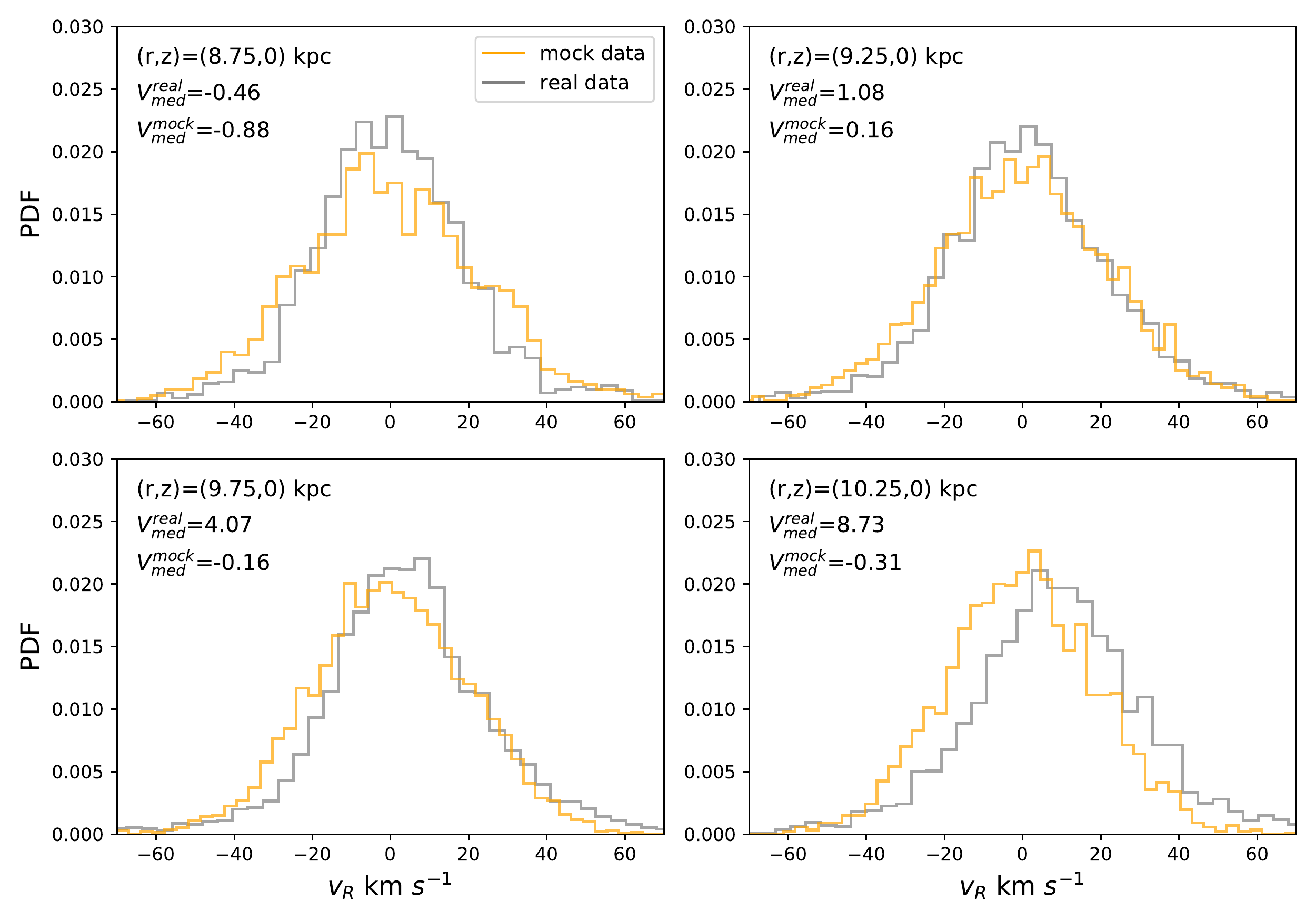}
    \caption{Curves: the distributions of $v_R$  predicted by the best-fit
    model in the plane at $R=8.75,\,9.25,\,9.75,$ and $10.25\kpc$.
 Histograms: distributions of $v_R$ for the OB stars in the LAMOST catalogue
 that lie in these spatial bins.
The number of stars in each bin is 2097, 3572, 4268, and
3215.}\label{fig:vdf}
\end{figure*}

Fig.~\ref{fig:Vlos} shows the distributions of the corrected values of
$v_\parallel$ when the stars are divided into four bins in $R$ with the
distributions of $v_\parallel$ obtained by sampling the best-fit model at the
locations of observed stars. The observed and mock distributions have similar
shapes but are offset from one another by an amount that grows with $R-R_0$.
In principle  these offsets could arise through  biased distances for
the stars causing Galactic rotation to make erroneous contributions to the
mock values of $v_\parallel$. We found that systematically increasing
distances by $10$ per cent  shifted even the  mock histogram for the furthest
bin by only $\sim1.5\kms$, so neither erroneous distances nor faulty data
reduction can explain the measured
offsets.

Fig.~\ref{fig:vdf} compares observed and mock distributions of $v_R$ at the
locations of LAMOST stars. This comparison is more sensitive to adopted
distances than the above comparison of $v_\parallel$ values because $v_R$ has
contributions from $v_\ell$ and $v_b$, which are proportional to distance.
These contributions are small, however, because all stars lie near the
anticentre. The mock distributions from our axisymmetric model are inevitably
symmetric in $v_R$, so the comparison highlights the asymmetry in the
observed distributions. The asymmetry is negligible for the nearest sample
because the $U$ component of the solar velocity is chosen to eliminate this
asymmetry in the wider populations of disc stars. The observed systematic
increase in asymmetry with $R-R_0$ must arise from some combination of spiral
structure and large-scale distortion of the disc by the Sgr Dwarf galaxy,
which has been extensively discussed in connection with the Galactic warp and
the phase spiral
\citep{JiangBinney,AntojaSpiral,Laporte2019,BinneySchoenrich2018,BlandHawthorn2019}.
Fig.~\ref{fig:vdf} is consistent with what one would expect from the map ov
$v_R$ deduced by \cite{Eilers2020} from data for red giant stars.

\begin{figure}
	\includegraphics[width=\columnwidth]{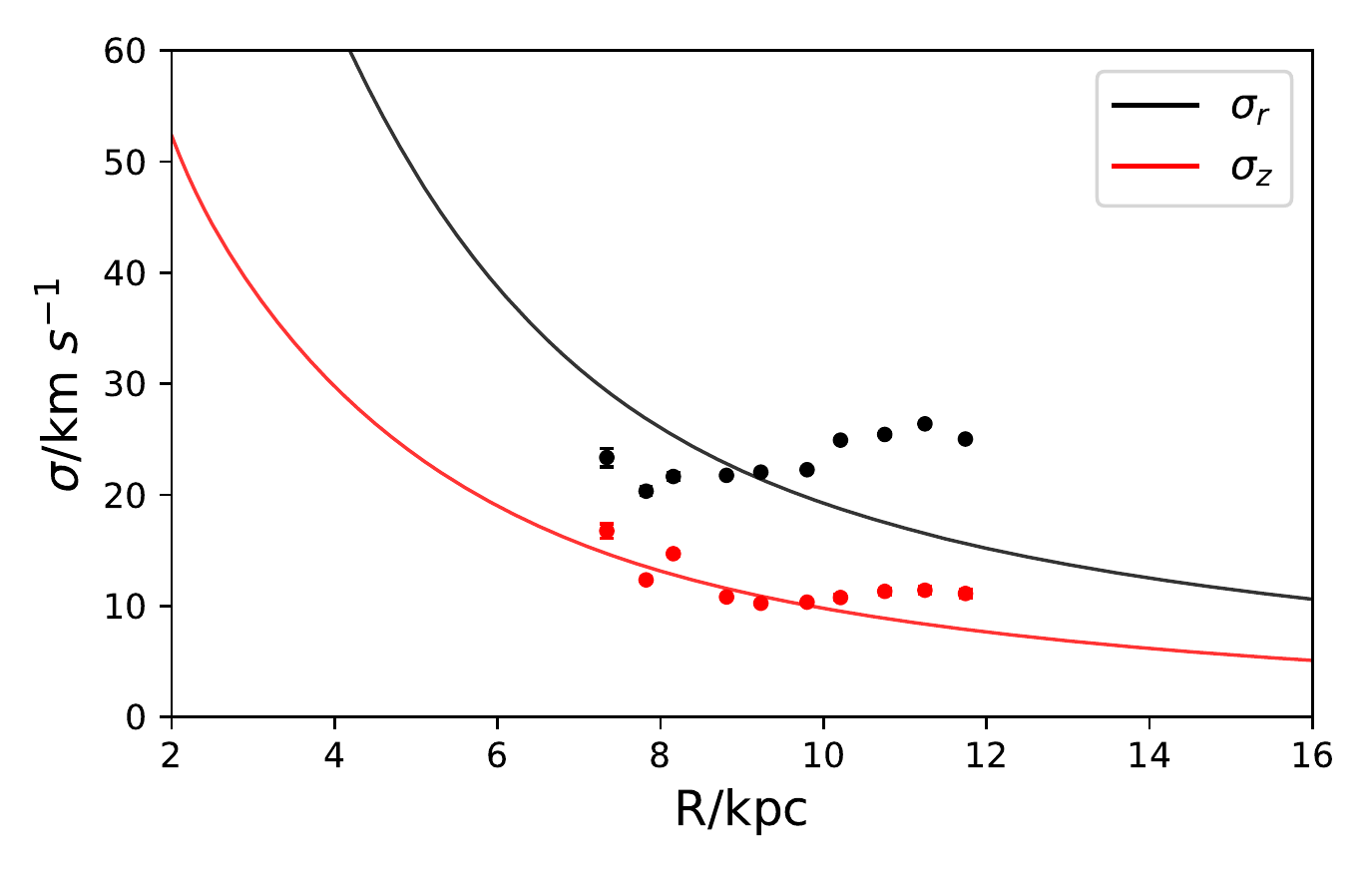}
    \caption{The curves show the profiles of the velocity dispersions $\sigma_{z}$ (red) and
$\sigma_{R}$ (black) predicted by the best-fit model. The
dots show values derived from OB stars in LAMOST.}
    \label{fig:dispersion_lamost}
\end{figure}

\begin{figure}
	\includegraphics[width=\columnwidth]{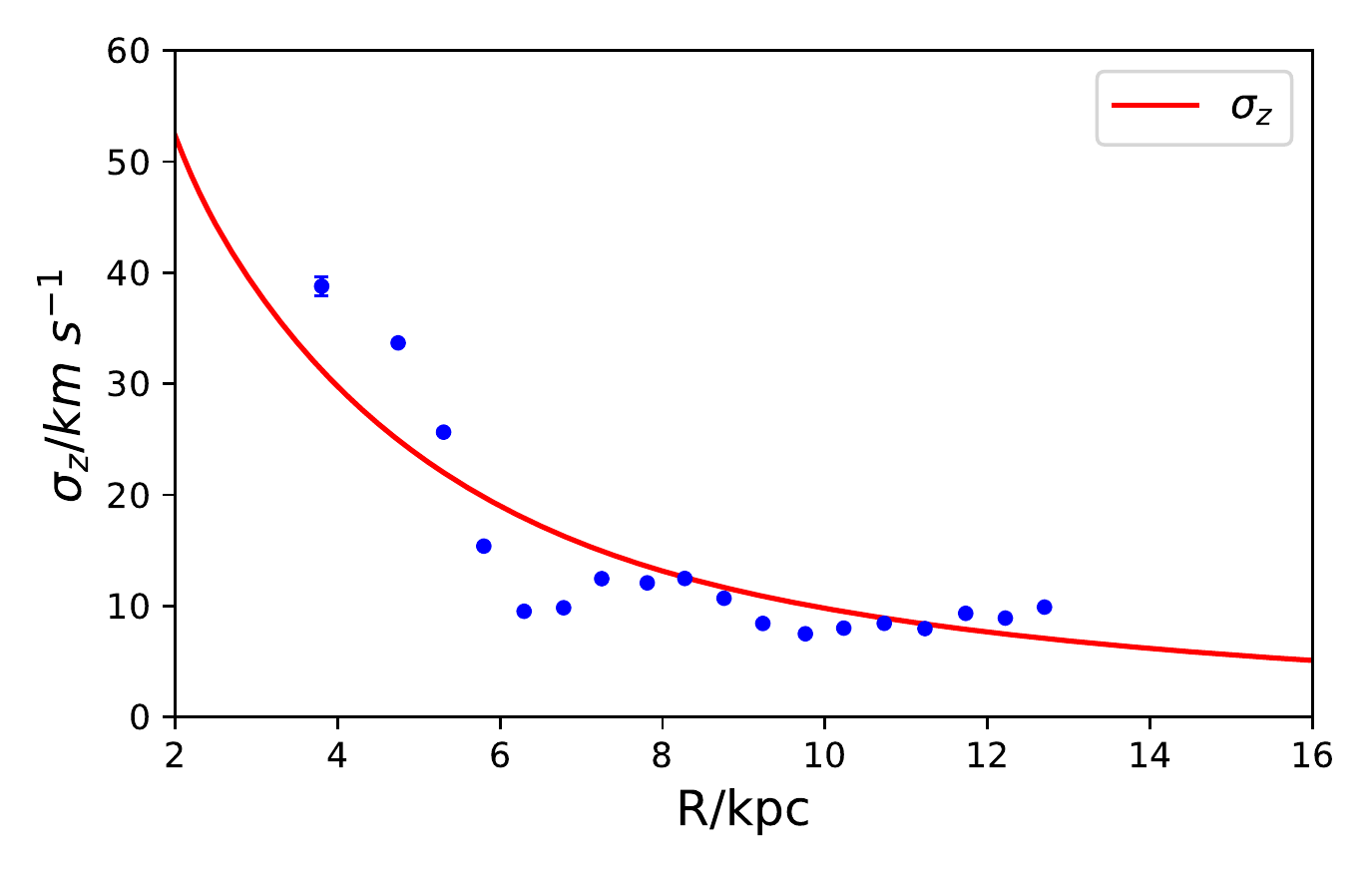}
    \caption{The curve shows $\sigma_z$ as a function of  $R$ in the
    Galactic plane from the best-fit model. The dots show $v_b$ for real
    stars in Gaia EDR3 with $|b|<5\,$deg.}
    \label{fig:vdis_real}
\end{figure}

Fig.\ref{fig:dispersion_lamost} compares our model's predictions (curves) for
the velocity dispersions $\sigma_z$ and $\sigma_R$
with results from OB stars in LAMOST. The data for $\sigma_z$ (red line and
points) are in reasonable concordance, although the LAMOST data show a
weaker outward decline than the model predicts.  Fig.~\ref{fig:vdis_real} provides
evidence that the steeper gradient of the model is required by the EDR3 data
by comparing the model prediction (dashed line) with the dispersion in $v_b$,
which for these low-latitude stars differs little from $v_z$.

\begin{figure}
	\includegraphics[width=\columnwidth]{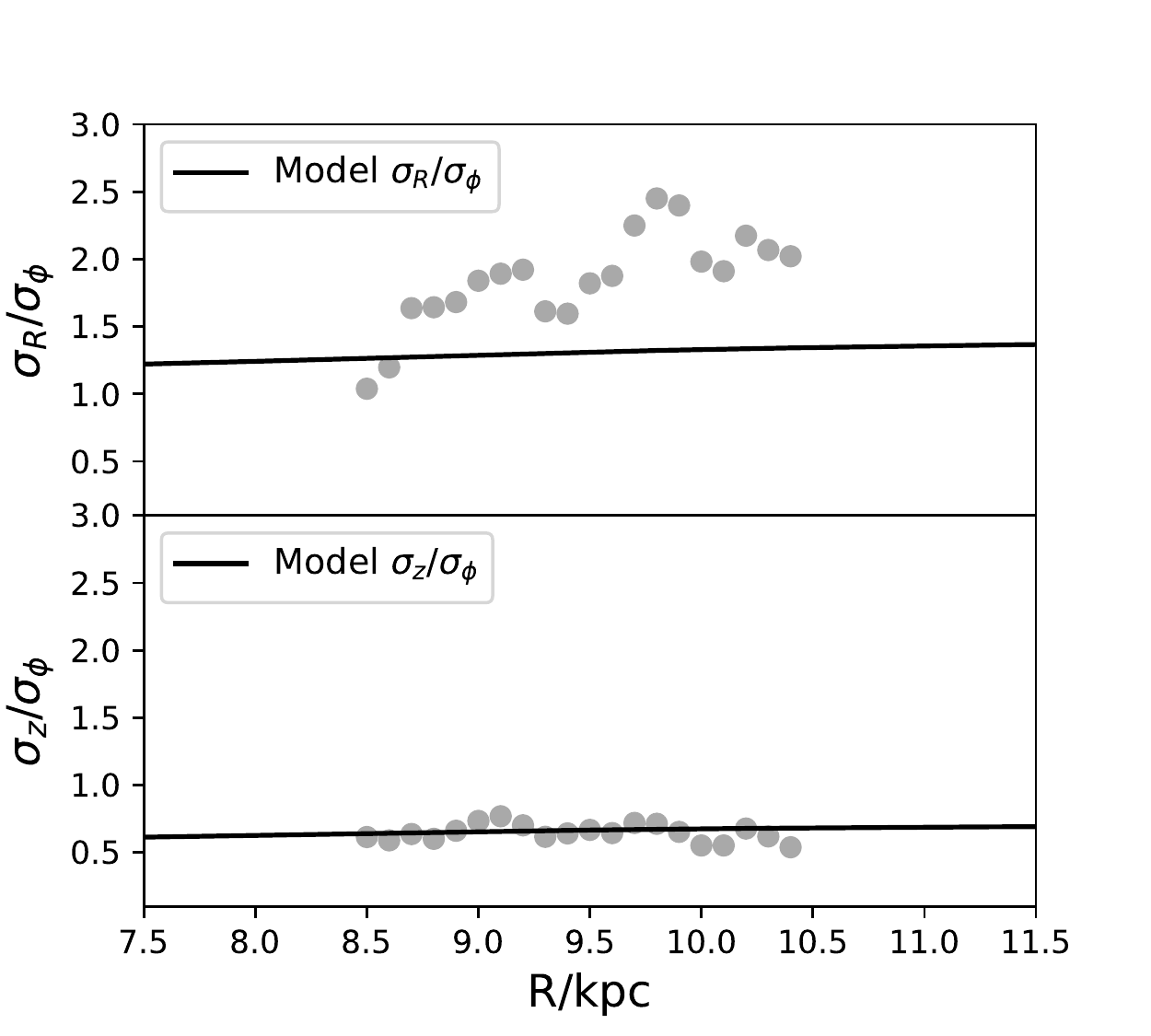}
    \caption{$\sigma_R/\sigma_\phi$ and $\sigma_z/\sigma_\phi$ plotted against $R$ for OB stars in
    LAMOST using Starhorse distances (right panel) or these distances multiplied by
    $1.1$ (left panel).}
    \label{fig:sigmaratio}
\end{figure}

In Fig.~\ref{fig:dispersion_lamost} the black points and full curve for
$\sigma_R$ are starkly incompatible. This conflict signals that the $v_R$
distribution has been broadened as well as shifted to negative values. The
points in the upper panel of Fig.~\ref{fig:sigmaratio} show the ratio
$\sigma_R/\sigma_\phi$ as a function of $R$ for LAMOST stars, while the black
line shows the prediction of the best-fit model.  The data points climb away
from the model's line as $R-R_0$ grows.  A classical result of stellar dynamics
is that $\sigma_R/\sigma_\phi$ is largely set by the shape of the circular-speed curve:
when the latter is flat, any thin-disc population will have
$\sigma_R/\sigma_\phi\simeq\surd2$ \citep[e.g.][]{GDII}.  The LAMOST data imply
values of $\sigma_R/\sigma_\phi$ that reach well above 2. No equilibrium
model could match these values.

\subsection{Impact of binaries}\label{sec:binarity}

We now ask whether binary stars can explain the anomalously large values of
$\sigma_R/\sigma_\phi$ plotted in Fig.~\ref{fig:sigmaratio}.

Massive stars usually have a massive binary companion, but we find that
only $\sim$10 per cent of the OB stars in our sample have binary companions
resolved by Gaia. Taking the angular resolution of Gaia to be $0.1\,$arcsec,
it follows that the majority of these stars must be in binaries with
separations
\[\label{eq:bin_sep}
r<0.2{s\over\hbox{pc}}\,\hbox{AU}.
\]
Consequently their orbital velocities satisfy
\[\label{eq:bin_v}
v_{\rm b}^2>45\bigg({M_1+M_2\over10\msun}{\hbox{kpc}\over s}\bigg)(\!\kms)^2.
\]
The least massive B star has $M\sim3.8\msun$ \citep{GA}, so binary velocities
in excess of $10\kms$ must be common. 

The velocities of binaries will be isotropically distributed, so the
mean-square value of the component along the line of sight is
$v_{\parallel\rm b}^2=v_{\rm b}^2/3$. The mean-square velocity of the primary
is $M_2^2/(M_1+M_2)^2$ times this, so smaller by a factor of at least 4.  If
for simplicity we assume that light from the primary dominates the spectrum,
the contribution to the measured value of $v_\parallel^2$ is
\[
\begin{aligned}
v_{\parallel1}^2&=\fracj{1}3{GM_\odot\over1\,\hbox{AU}}
\bigg\langle{1\,\hbox{AU}\over r}\bigg\rangle{M_1+M_2\over
\msun}\Big({M_2\over M_1+M_2}\Big)^2\cr
&\simeq300\bigg\langle{1\,\hbox{AU}\over r}\bigg\rangle
{M_2\over M_1+M_2}{M_2\over M_\odot}(\kms)^2
\end{aligned}
\]
The distribution of binary separations determined by \cite{Duchene2013} and \cite{GravityCollaboration2018} in a study of
the Orion nebula yields $\big\langle
1/r\big\rangle\simeq5\,\hbox{AU}^{-1}$, so
\[
v_{\parallel1}^2\simeq(39)^2{M_2\over M_1+M_2}{M_2\over M_\odot}(\kms)^2
\]
Given that the product of masses in this equation is of order unity and that
$\sigma_R$ is dominated by $v_\parallel$, it is very much to be expected that
binaries will cause the measured value of $\sigma_R/\sigma_\phi$ to be
substantially higher than expected in a model that ignores binaries.
Fig.~\ref{fig:dispersion_lamost} suggests that binaries set a floor value,
$\sigma_R\simeq20\kms$, to $\sigma_R$, which is smaller by a factor $\sim2$
than the value suggested by the calculation above. Our value of
$\langle1/r\rangle$ may well be too large because it is dominated by the
small fraction of very close binaries, and is correspondingly uncertain.

\section{Conclusions}\label{sec:conclude}

We extracted a sample of $\sim47\,000$ OB stars from the intersection of Gaia
EDR3 with the 2MASS and Pan-STARRS surveys. For this sample we have photometry
plus five-dimensional astrometry. We used the algorithm developed in
\cite{Li2022} in connection with similar data for a sample of RR-Lyrae stars
to fit DFs of the form $f(\vJ)$ to the OB population. Since the surface
density of the young disc is expected to peak at a radius of a few kpc, we
introduced two new parameters to the disc DFs proposed by BV2022. Small
extensions of the algorithm were required to deal with (i) the finite spread
in luminosity of the sampled stars, and (ii) extinction by dust. Tests of the
algorithm on similar pseudo-data showed that the parameters of the true DF
can be recovered with remarkable precision when the distribution of dust is
accurately known. When a significantly incorrect dust distribution is used to
analyse the data, the true parameters can lie several sigmas above or below
the probable range returned by the algorithm.

When the algorithm is run on the real data using the dust model of
\cite{Green2019}, the most probable DF yields pseudo-data that are not
correctly  distributed on the sky. The source of this discrepancy is almost
certainly imperfections of the dust model used. Since
the distribution of velocities depends quite strongly on location,
discrepancies in the spatial distribution of stars automatically leads to
discrepancies between the predicted and measured velocity distributions. When
the most probable DF is used to sample velocities at the observed locations
of stars rather than at the locations of pseudo-stars, the observed and
predicted velocity distributions agree well. These tests show that a sample of OB
stars like the present one would pin down the phase-space structure of the
young stellar disc with impressive accuracy if the three-dimensional
distribution of dust were accurately known. The discrepancies between the
distributions of pseudo-stars and real stars constitute clear evidence that
the best current dust model is significantly flawed.

Although the functional form of the DF provides great flexibility in the
radial distribution of stars, the best-fit DF predicts a distribution of OB
stars that extends to smaller radii than the observed sample. Since the
kinematics of a stellar disc are not independent of its surface-density
profile, the MCMC search may favour discs that extend unexpectedly far in
because the kinematics of such discs may fit the data better. Moreover, a dust
model that included more  dust interior to the Sun would make the data
consistent with more extended discs. Hence, we consider it likely that there
is more dust at $R\la 7\kpc$ than the \textit{Bayestar2019} model envisages.

The most probable DF predicts the distributions of the space velocities of OB
stars. We have compared these predictions with the distributions of the
sub-sample of EDR3 OB stars that have measured line-of-sight velocities
because they fell within the LAMOST survey. The predicted and measured
dispersions $\sigma_z$ agree fairly well, although the measured values of
$\sigma_z$ are flat beyond $R\sim10\kpc$ while the predicted values decrease
monotonically outwards. The measured values of $\sigma_R$ do not decrease
outwards as the model predicts -- they even increase slightly. Moreover, the
median value of $v_R$ drifts upwards from near zero at $R_0$ to
$\overline{v}_R\sim10\kms$ at $R\sim12\kpc$.  We concluded that this trend in
$\overline{v}_R$ is not an artifact induced by erroneous distances. In
particular, it is associated with similar offsets between the measured
line-of-sight velocities and those predicted by the model. It seems that some
combination of spiral structure and disturbance by the gravitational field of
the Sgr dwarf galaxy is changing velocities by $\ga10\kms$ between here and
$R\sim12\kpc$.

The ratio $\sigma_R/\sigma_\phi$, which is
strongly constrained by the shape of the circular-speed curve, is
unexpectedly large for the LAMOST stars. We argued that this result arises
naturally from the majority of the stars being in binaries too tight for Gaia
to resolve. The line-of-sight velocity of such a star will differ from its
barycentric velocity by a fraction of the binary velocity because the LAMOST
data for most stars are based on a single epoch, The EDR3 proper motion, by
contrast, is obtained by fitting a curve through the star's observed
positions at $\sim40$ epochs, so will be barely affected by their binary
nature. The line-of-sight velocity of distant LAMOST OB stars feeds strongly
into $v_R$ rather than $v_\phi$ because they are located close towards the
anticentre. Hence binaries will push $\sigma_R/\sigma_\phi$ above the value
predicted for single stars.

The key to improving our understanding of the Galaxy's young disc is
construction of a better map of the distribution of dust. Traditionally dust
is mapped by measuring extinctions for stars with known distances and the
advent of vast number of precise parallaxes for stars within a few
kiloparsecs of the Sun has breathed new life into this field
\citep{BailerJones2011,Sale2014,Green2019,Lallement2022}. The major problems now are (i) the
determination of extinctions for large numbers of stars, and (ii)
synthesising the extinction measures into a coherent picture of the dust
distribution. This work suggests an approach to dust mapping that dispenses
with extinctions measures of individual stars. The stars of a stellar
populations that is more than $\sim200\Myr$ old have to be fairly smoothly
distributed in phase space.  Moreover, the population's velocity distribution
at one location strongly constrains the population's DF, and therefore its
spatial distribution.  Moreover, the velocity distribution at a location can
be determined without knowledge of the column of dust through which we see
that location because obscuration is velocity-independent. Hence, an exercise
along the present lines could yield fairly firm {\it predictions} for the
spatial distribution of stars, and a dust model could be strongly constrained
by comparing this predicted distribution in
$(\ell,b,\varpi,\mu_\ell,\mu_b,G,\hbox{colours})$ with the observed star
density.

A great advantage of such an approach is that it could exploit all the
$>1.3\,$billion stars tracked by Gaia, and thus produce dust maps of
unprecedented spatial resolution. To implement this idea, it is necessary
both to confront the computational challenge of simultaneously exploring
high-dimensional parameter spaces of DFs and dust models, and to obtain a good
model of Gaia's selection function. There is an excellent prospect that
implementation will soon be feasible.
 
\section*{Acknowledgements}
We thank the anonymous referee for
the suggestions to improve this paper. CL and JB are supported by the UK Science
and Technology Facilities Council under grant number ST/N000919/1. JB also
acknowledges support from the
Leverhulme Trust through an Emeritus Fellowship. 

This work presents results from the European Space Agency (ESA) space mission Gaia. Gaia data are being processed by the Gaia Data Processing and Analysis Consortium (DPAC). Funding for the DPAC is provided by national institutions, in particular the institutions participating in the Gaia MultiLateral Agreement (MLA). The Gaia mission website is https://www.cosmos.esa.int/gaia. The Gaia archive website is https://archives.esac.esa.int/gaia.

This publication makes use of data products from the Two Micron All Sky Survey, which is a joint project of the University of Massachusetts and the Infrared Processing and Analysis Center/California Institute of Technology, funded by the National Aeronautics and Space Administration and the National Science Foundation.

\section*{DATA AVAILABILITY}
The \AGAMA\, source codes and model parameter file are available  in the online supplementary material. The data underlying this article will be shared on reasonable request to the corresponding author.

\bibliographystyle{mnras}
\bibliography{example} 

\appendix
\section{Formulae for computing likelihoods}\label{app:LiB}

Let $f(\vw)$ be the DF normalised such that $\int\d^6\vw\,f=1$, and $\Phi(M)$
be the population's luminosity function.  Then the probability that a
randomly chosen survey star is located within the phase-space volume
$\d^6\vw$ around the phase-space location $\vw=(\vx,\vv)$ and has absolute
magnitude in $(M,M+\d M)$ is
\[
\begin{aligned}
P(\vw,M|f\hbox{ \& in survey})&\,\d^6\vw\,\d M
=\cr
&{S(\vw,M)f(\vw)\Phi(M)\over P_{\rm S}}\,\d^6\vw\,\d M,
\end{aligned}
\]
where $S(\vw,M)$ is the probability that a star at $\vw$ of absolute
magnitude $M$ enters the survey.  The denominator
\[
P_{\rm S}=\int\d^6\vw\,f(\vw)\int\d\vM\,\Phi(M)S(\vw,M)
\]
is the probability that a star randomly chosen from the population will
appear in the catalogue. It ensures that $P(\vw|f\hbox{ \& in survey})$ is a
correctly normalised probability density. 

On account of observational errors, we should maximise not $P(\vw,M|f\hbox{ \&
in survey})$ but the related probability density
\[
P_{\rm o}(\vw,M)=\int\d^6\vw'\,G(\vw-\vw',\vK)P(\vw',M|f\hbox{ \& in survey})
\]
that the catalogue will list a star of absolute magnitude $M$ at $\vw$ that
has true phase-space location $\vw'$. Here we assume that the distribution of
observational errors $G$ is a multi-variate Gaussian with kernel $\vK$:
\[
G(\vw,\vK)=\sqrt{|\vK|\over(2\pi)^n}\exp(-\fracj12\vw^T\cdot\vK\cdot\vw),
\]
where $n=\hbox{dim}(\vw)$.
Thus $f$ should be chosen to maximise
\[
P_{\rm o}(\vw,M)={1\over P_{\rm S}}\int\d^6\vw'\,G(\vw-\vw',\vK)
S(\vw',M)f(\vw')\Phi(M).
\]

In practice it is convenient to work with sky coordinates, which do not
comprise a system of canonical coordinates for phase space. Specifically, we
use Galactic longitude and latitude $\ell,b$, distance $s$, the proper
motions $\mu_\ell=\dot\ell\cos b$ and $\mu_b=\dot b$, the line-of-sight velocity $v_\parallel$. 
Forming these into the vector $\vu=(\ell,b,s,\mu_\ell,\mu_b,v_\parallel)$, we
have that the element of phase-space volume $\d^6\vw$ is related to $\d^6\vu$
by \citep[e.g][]{McMillan2012}
\[
\d^6\vw=s^4\cos b\,\d^6\vu,
\]
so
\[\begin{aligned}
P_{\rm o}(\vu,M)&={1\over P_{\rm S}}\int\d^6\vu'\,s^{\prime 4}\cos
b'\,G(\vu-\vu',\vK)\cr
&\hskip1cm\times S(\vw',M)f(\vw')\Phi(M),
\end{aligned}
\]
where $\vw'$ is understood to be a function of $\vu'$. One advantage of
working with sky coordinates is that the matrix $\vK$ then simplifies. Most
of its off-diagonal elements vanish and $K_{\ell\ell}$ and $K_{bb}$ become
very large because the sky positions of stars have negligible uncertainty.
This being so $G$ may be approximated by the product of a $4\times4$ matrix
$\widetilde\vK$ and Dirac $\delta$-functions in $\ell-\ell'$ and $b-b'$ so
the integrals over these coordinates can be trivially executed. Otherwise we neglect correlations by
approximating $\widetilde\vK$ by a diagonal matrix,
\[
\widetilde\vK=\hbox{diag}\big[\sigma_s^{-2},\sigma_{\mu_\ell}^{-2},
\sigma_{\mu_b}^{-2},\sigma_{v_\parallel}^{-2}\big].
\]

Conversion of heliocentric coordinates into phase-space coordinates requires
knowledge of the Sun's Galactocentric position and velocity.  We use
Galactocentric Cartesian coordinates $(x,y,z)$ in which the Sun's position
vector is $ (-8.2, 0, 0)$. Heliocentric distances are denoted by $s$ and
Galactocentric distances by $r$.  From \cite{Schoenrich2012} we take the Sun's
Galactocentric velocity $\vV_\odot$ to be
\[
\vV_\odot=(U,V,W)=(11.1, 250.24, 7.25)\kms.
\]

In practice $S$ depends not on $M$ but on the apparent magnitude $m=M+\mu+A$, where
$\mu=5\log_{10}(s/10\pc)$ is the distance modulus and $A(\vx)$ is the extinction.

\subsection{Evaluating the probabilities}\label{sec:compute}

Evaluation of the quality of a model requires execution of two distinct
numerical tasks. One is computation of $P_{\rm S}$  by integrating the
product of the DF and the survey's selection function through phase space.
Introducing angle-action variables $(\vtheta,\vJ)$,
we have
\[
P_{\rm S}=\int\d^3\vJ\,f(\vJ)\int\d^3\vtheta\int\d M\,S(\vJ,\vtheta,M)\Phi(M).
\]
Following \cite{McMillan2013} we execute this integral by the Monte-Carlo
principle:
\[
\int \d x\,f(x)\simeq{1\over N}\sum_{i=1}^N{f(x_i)\over f_{\rm s}(x_i)},
\]
where the points $x_i$ are randomly sampled according to the probability
density $f_{\rm s}$. We take $f_{\rm s}$ to be a function of $\vJ$ only, so
we can write
\[
P_{\rm S}\simeq{1\over N}\sum_i^N{f(\vJ_i)\over f_{\rm
s}(\vJ_i)}\,\int \d M\,\Phi(M)\,S(\vtheta_i,\vJ_i,M).
\]
Poisson noise is minimised if $f(\vJ)/f_{\rm s}(\vJ)\simeq1$, and in our case
this can be achieved by taking $f_{\rm s}$ to be the first DF we try. If the
coordinates $(\vtheta_i,\vJ_i)$ of the points that sample $f_{\rm s}$ and the
resulting values of the ratio $S(\vtheta_i,\vJ_i)/f_{\rm s}(\vJ_i)$ are
stored, the quality of any subsequently proposed DF can be computed cheaply
merely by evaluating it at the $\vJ_i$.

We use Gauss-Legendre integration with $N=20$ nodes to evaluate the second integral for a k-th star:
\begin{equation}\label{eq:pin_sec}
    \int_{M_{k}^{-}}^{M_{k}^{+}} \d M\,\Phi(M)\,S(\vx_k,M)
    =\frac{M_{k}^{+}-M_{k}^{-}}{2}\sum_{i=1}^{20} A_i g(M_i),
\end{equation}
where $\vx$ are fixed positions given the sampling density and $M_{k}^{+}$
and $M_{k}^{-}$ are, respectively, the upper and lower limit of the absolute magnitude
respectively:
\begin{equation}\label{eq:abolim}
    \begin{aligned}
    &M_{k}^{+}~=~m_{k}^{+}\,-5\times\log_{10}{\frac{s_{k}}{1\pc}}\,-A_{k}\\
    &M_{k}^{-}~=~m_{k}^{-}\,-5\times\log_{10}{\frac{s_{k}}{1\pc}}\,-A_{k},
    \end{aligned}
\end{equation}
where $m_{k}^{+}=21$ and $m_{k}^{-}=5$ are the apparent magnitude limits of
the
Gaia survey, $A_k$ is the  G-band extinction. Note that these are the
astrometric solutions' limits rather than the detection limits of the Gaia
survey. Additionally, if the computed absolute magnitude limits exceed the
boundaries in the LF which is shown in Figure~\ref{fig:OBLF}, the boundary
value will then be used to replace the computed absolute magnitude limits in
the integral.  Since we ignore the colour criteria in the SF, we only need to
deal with LF in Gaia G band in the normalization factor.

We are concerned with the case that $v_\parallel$ has not been measured. Then
the error ellipsoid becomes a section of a four-dimensional cylinder, the
cross-sections of which are three-dimensional ellipsoids spanned by the
measured quantities $(s,\mu_\ell,\mu_b)$.  We need
to integrate through only that part of the cylinder for which the
Galactocentric speed $v$ is less than the escape speed because the DF
vanishes for $v(v_\parallel)>v_{\rm esc}$. The strategy we adopted is to
obtain from {\tt quadpy}\footnote{https://github.com/nschloe/quadpy} $n=77$
locations $\vx_i$ and weights $A_i$ for three-dimensional integration with
weight function $\e^{-|\vx|^2}$. Then with $G(\vu|\vK)$ now denoting a
three-dimensional Gaussian distribution, we have at each $i$ that
\begin{align}
P_{\rm o}
&={\cos b\over V_0P_{\rm S}}
\int\d^3\vu\,G(\vu-\vu'|\vK)\int_{v_-}^{v_+}\d v_\parallel\,h(\vu,v_\parallel)\cr
&={\cos b\over V_0\pi^{3/2}P_{\rm S}}
\sum_{n=1}^{77} A_i \int_{v_-}^{v_+}\d v_\parallel\,h_i(v_\parallel),
\end{align}
where $V_0$ is a normalising velocity, $v_\pm$ are the values of $v_\parallel$  at which $v=v_{\rm esc}$ and 
\[
h_i(v_\parallel)=h\Big(\vu_i+\surd2\!\!\sum_{\alpha=\ell,b,s}\!\!\sigma_\alpha x_{i\alpha}\ve_\alpha
+v_\parallel\ve_\parallel\Big).
\]
The integral over $v_\parallel$ was executed by Gauss-Legendre integration
with unit weight function using $35$ integrand evaluations.\footnote{The
overall count of $77\times35=2695$ evaluations of the DF for each star.} The
number of node for disc stars are more than 1.5 times that in the RR-Lyrae
work, which is because the line-of-sight distribution for disc stars are not
regularly due to the SF and extinction on the plane.  The constant $V_0$ is
the same for all stars and DFs, so it plays no role in the optimisation and
can be set to unity.

\section{Test result for wrong dust model}\label{sec:testfalsedust}

\begin{figure*}
	\includegraphics[scale=0.45]{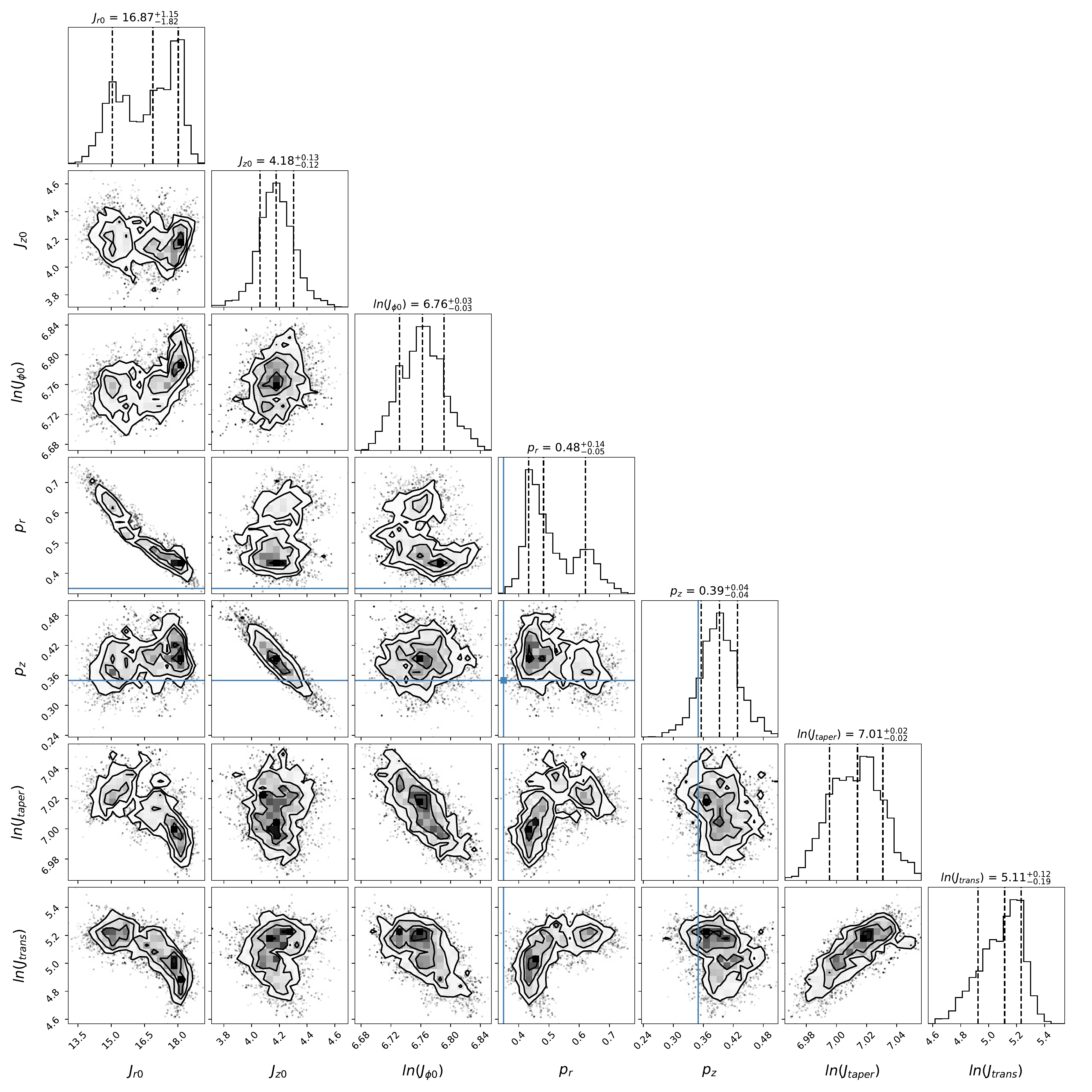}
    \caption{Posterior probability distributions obtained when a mock
    catalogue created using the toy dust model is analysed using the
    \textit{Bayestar2019} dust model.}
    \label{fig:mock_wrong_dust}
\end{figure*}

Fig.~\ref{fig:mock_wrong_dust} shows the posterior distributions when the
\cite{Green2019} dust model is used to analyse mock data produced using the
toy dust model of Section~\ref{sec:toyDust}.

\label{lastpage}
\end{document}